\documentclass{article}
\usepackage[utf8]{inputenc}

\usepackage{natbib}
\setcitestyle{authoryear,open={(},close={)}}

\usepackage{amsmath,amssymb,amsfonts,mathtools}
\usepackage{graphicx}
\usepackage{fancyhdr}
\usepackage{hyperref}
\hypersetup{
    colorlinks = true,
    urlcolor   = blue,
    citecolor  = black,
}
\usepackage{geometry}
\newcommand{\that}{\hat{\boldsymbol{t}}}
\newcommand{\rhat}{\hat{\boldsymbol{r}}}
\newcommand{\muhat}{\hat{\boldsymbol{\mu}}}
\newcommand{\nuhat}{\hat{\boldsymbol{\nu}}}
\newcommand{\ehat}{\hat{\boldsymbol{e}}}
\newcommand{\bU}{\boldsymbol{U}}
\newcommand{\bX}{\boldsymbol{X}}
\newcommand{\bY}{\boldsymbol{Y}}
\newcommand{\bLambda}{\boldsymbol{\Lambda}}
\newcommand{\br}{\boldsymbol{r}}
\newcommand{\bcdot}{\boldsymbol{\cdot}}
\newcommand{\bnabla}{\boldsymbol{\nabla}}
\newcommand{\vect}{\textbf{vec}}
\newcommand{\idvec}{\mathbb{I}}

\begin{document}
\begin{center}
    {\Large\bfseries A Comparison of Instabilities and Dynamic States in Active Filament Models}\\[2ex]
    {\large Ilteber R. Ozdemir\textsuperscript{1}, Bethany Clarke\textsuperscript{1}, Yongyun Hwang\textsuperscript{2} and Eric E. Keaveny\textsuperscript{1$\dagger$}}\\[1ex]
    {\small
        \textsuperscript{1}Department of Mathematics, Imperial College London, SW7 2AZ, UK\\
        \textsuperscript{2}Department of Aeronautics, Imperial College London, SW7 2AZ, UK
    }\\[2ex]
\end{center}
\begingroup
\renewcommand{\thefootnote}{$\dagger$}
\footnotetext[0]{\centering Email address for correspondence: e.keaveny@imperial.ac.uk}
\endgroup

Active filaments, such as microtubules with attached cargo-carrying motor proteins, are important dynamic structures for fluid transport in and around living cells. The mathematical models of active filaments appearing in the literature typically involve combinations of follower-forces, compressive tangential forces, along the filament, and an opposite force on the fluid that generates an effective surface flow. In this paper, we present a comparative dynamical systems study of active filament models examining the differences in dynamic states that occur when actuation is through follower forces alone, or the effect of surface flows is also included. We consider cases where actuation is applied only at the filament tip, or distributed uniformly along the filament length. By varying actuation strength, we show that the first bifurcations that provide the transition between the upright, whirling and beating states appear in all models. At higher values of actuation, when beating becomes unstable, however, qualitative differences between the models emerge. Those with distributed actuation produce a single, time-dependent state, which for the surface flow model is reminiscent of a rotating helix that periodically changes handedness and rotation direction. Tip actuation, however, yields complex transitions that ultimately produce a chaotic state. We link the differences in dynamics between tip and distributed actuation to differences in their respective internal stress distributions, differences which appear as early as the first bifurcation where they affect the shapes of the unstable modes.

\section{Introduction}\label{sec:introduction}
Living cells generate fluid motion in and around them to facilitate fundamental biological processes such as motility, sensing, and feeding. These fluid flows are essential to overcome the limitations of molecular diffusion for mass transport, and further, they must be effective in a reversible, viscosity-dominated environment where the effects of inertia are negligible \citep{lighthill1976,lauga2020}. One way of creating these flows is through cargo-carrying motor proteins that move along biological filaments, so-called active filaments \citep{mallik2004,cho2012}. Within the cell, motor proteins move along microtubules, generating effective surface velocities by pushing the surrounding fluid as they move. Interactions between neighbouring active filaments enable coordinated fluid flows within the cell, which are critical, for example, to the proper development of fruit fly eggs \citep{stein2021,htet2023cortex,dutta2024}, and to the control and positioning of metaphase spindles during cell division \citep{wu2023,young2025}. Active filaments also comprise the axoneme, the internal structure of cilia, where motor proteins generate the internal shear that produces cilium beating. Cilia are responsible for fluid motion outside cells in processes that include the locomotion of microscopic organisms and cells \citep{berner1993,gaffney2011}, and the pumping of fluids in the vital organs of more complex organisms \citep{sleigh1988,lyons2006,faubel2016}.

Experiments suggest that the pico-Newton scale active forces exerted by the motor proteins are sufficient to deform the biofilament along which they move \citep{svoboda1994}. As a result, active filament motion arises from the fluid-structure interactions driven by the stresses associated with motor protein activity. Since the Reynolds number associated with filament motion is low, models for active filaments can take advantage of theoretical and computational methods for slender, flexible bodies in Stokes flow. These often couple either Euler-Bernoulli or Kirchoff rod theories with hydrodynamic models based on local resistive force theory \citep{gray1955,lighthill1976} and nonlocal slender-body theory \citep{johnson1980,tornberg2004,nazockdast2017fast}, or instead methods based around regularised Stokeslets, Rotne-Prager Yamakawa tensor, immersed-boundary and force-coupling methods \citep{maxian2021integral,maxian2022hydrodynamics,fuchter2023three,hall2019efficient,jabbarzadeh2020numerical,delmotte2015general,olson2013modeling,lim2010dynamics,walker2020efficient,schoeller2021} that resolve hydrodynamic interactions between different elements of the filament. While these approaches provide descriptions of the passive elastic and viscous forces that are present, the active forces that represent motor activity and drive filament motion must also be incorporated.

A first approach to incorporating motor activity is to include a compressive, tangential load \citep{beck1952,herrmann1964,langthjem2000}, so-called follower force, to represent the forces that the motors exert on the filament. Using resistive force theory, \cite{decanio2017} considered a single follower force at the free-end of a filament that is clamped at its other end, and analysed its planar dynamics as the magnitude of the follower force increases. They showed that at a critical load, the filament buckles and undergoes a supercritical Hopf bifurcation leading to a time-periodic beating state. Allowing for fully 3D dynamics, \cite{ling2018} instead observed a whirling state just after buckling and planar beating arises at higher values of the follower force. They showed also how these states can change depending on where the follower forces are placed along the filament. To better understand the differences between the planar and fully-3D results, \cite{clarke2024} performed a computational dynamical systems analysis of the tip driven model and found that at buckling both the planar beating and whirling solutions emerge through a double Hopf bifurcation, with whirling being stable. This scenario was later confirmed by \cite{schnitzer2025} who performed a weakly nonlinear analysis around the bifurcation. Additionally, \cite{clarke2024} found that a stable quasiperiodic state provides the transition from whirling to beating that occurs at higher follower forces. At even higher follower force values, beating is found to be unstable and various quasiperiodic and chaotic states are observed instead \citep{ling2018,krishnamurthy2023,clarke2024}.

In addition to follower forces acting on the filament, the motor proteins will exert the opposite force on the fluid, resulting in entrainment, or a surface flow, in the vicinity of the filament. The additional effect of the surface flow has been highlighted in modelling the collective dynamics of microtubules involved in cytoplasmic streaming \citep{stein2021} and is essential to the onset of large-scale flows. Despite this, less is known about single filament dynamics when a surface flow is included. When it has been incorporated into the follower force model, surface flows have been shown to counter the compressive effects of the follower force and shift buckling to higher follower force values \citep{decanio2017}. In exploring a closely-related active colloidal filament model, where a spherical squirmer is at the tip of an otherwise passive filament, \cite{laskar2017} found several dynamic states, such as beating and whirling, qualitatively similar to those that arise in the tip-driven, follower force model.

In this paper, we present a comparative dynamical systems study to identify filament states and their bifurcations for models that include or ignore surface flows. We also consider cases where activity is localised to the filament's free-end, or distributed along the filament length. Our computations reveal that filament states and bifurcations in all models are qualitatively similar at low actuation, though the inclusion of surface flows shifts bifurcations to higher activity values, consistent with previous studies. At higher actuation, however, we find that it is the distribution of activity that plays a key role in the transitions and resulting states. Actuation distributed uniformly along the filament length leads to a single, time-dependent state that remains stable even at very high values of actuation. When activity is concentrated at the free-end, however, the filament undergoes a complex series of bifurcations culminating in chaotic behaviour. We attribute these differences in dynamic states at high actuation to the internal stress profiles that arise in the different models.

The paper is structured as follows. In \S\ref{sec:Methods}, we present the different active filament models. In this section, we also describe the numerical methods we use to solve the resulting equations, as well as the computational dynamical systems tools we employ to obtain different states, assess their stability, and analyse quasi-periodic solutions. In \S\ref{sec:ResultsandDiscussions}, we present the main results of this study where we discuss and compare the bifurcation diagrams and dynamic states emerging form the active filament models, dividing results between the low and high actuation regimes. Finally, in \S\ref{sec:Conclusions} we present the conclusions of our study.

\section{Methods}\label{sec:Methods}
\subsection{Active Filament Models}
\subsubsection{Force and Moment Balances}
We begin by presenting the modelling framework we use to describe filament dynamics when driven either by follower forces or surface flows. In the follower force model, the filament is subject to a distribution of compressive tangential forces and entrainment is ignored. Our implementation follows directly from \cite{clarke2024} and \cite{schoeller2021} where only a tip-driven follower force was considered. In the surface flow model, the effects of both compressive forces and entrainment are incorporated by introducing an effective surface flow based on the squirmer model \citep{lighthill1952,blake1971}, similar to the model considered by \cite{laskar2017}.

\begin{figure}
    \centering
    \includegraphics[width=0.80\textwidth]{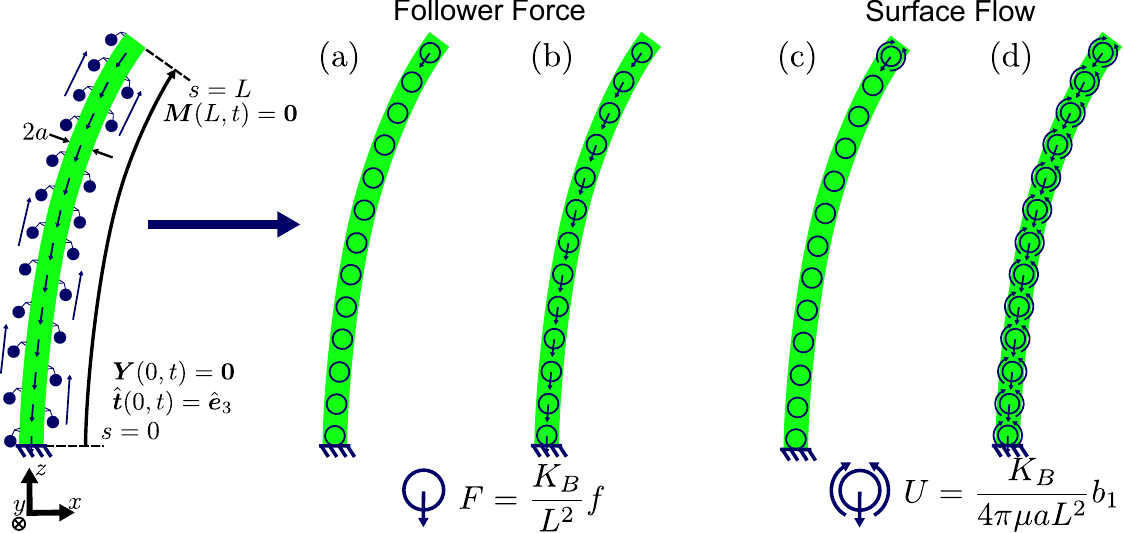}
    \caption{An illustration of the active filament models we consider: (a) tip-driven follower force, (b) distributed follower force, (c) tip-driven surface flow, and (d) distributed surface flow. The filament is clamped at its base in an unbounded domain and has a free end. $\ehat_1$, $\ehat_2$ and $\ehat_3$ are the unit vectors in $x$-, $y$- and $z$-directions.}
    \label{fig:filmodel}
\end{figure}

For both the follower force and surface flow models, we consider a filament whose base is held fixed at the origin and aligned with $\ehat_3$, as illustrated in Figure~\ref{fig:filmodel}. The basis vectors $\{\ehat_1$, $\ehat_2$, $\ehat_3\}$ provide the $x$-, $y$-, and $z$-directions, respectively, in the lab frame. The filament has length $L$ and cross-sectional radius $a$, and bending and twisting rigidities $K_B$ and $K_T$, respectively. In this work, we take $K_T = K_B$.  The filament is submerged in an unbounded fluid of viscosity $\eta$. The filament's centreline position is denoted by $\bY(s,t)$, where $s\in[0,L]$ is the arclength and $t\in[0,\infty)$ is time. We describe filament deformation using a local orthonormal basis, $\left\{\that(s,t),\muhat(s,t),\nuhat(s,t)\right\}$, at each material point along the centreline. Here, $\muhat(s,t)$ and $\nuhat(s,t)$ span the filament cross-section at $s$ while $\that(s,t)$ is constrained to be the unit-tangent vector through
\begin{equation}\label{eq:thatDef}
    \that(s,t)=\dfrac{\partial\bY(s,t)}{\partial s}.
\end{equation}

In the over-damped limit where inertia is negligible, the equations of motion are provided by the force and moment balances,
\begin{equation}\label{eq:contBalance}
\begin{aligned}
    \dfrac{\partial\bLambda}{\partial s}+\boldsymbol{f}^H+\boldsymbol{f}^A&=0,\\
    \dfrac{\partial\boldsymbol{M}}{\partial s}+\that\times\bLambda+\boldsymbol{\tau}^H&=0,
\end{aligned}
\end{equation}
where $\boldsymbol{f}$ and $\boldsymbol{\tau}$ denote force and torque (per unit length) acting on the filament, and superscripts $(\cdot)^H$ and $(\cdot)^A$ indicate the hydrodynamic and active contributions. The internal force and internal moment on the filament cross-section are denoted by $\bLambda(s,t)$ and $\boldsymbol{M}(s,t)$, respectively. The internal moment is given by the constitutive law \citep{landau1986},
\begin{equation}\label{eq:internalmoment}
    \boldsymbol{M}(s,t)=K_B\left(\that\times\dfrac{\partial\that}{\partial s}\right)+K_T\left(\nuhat\bcdot\dfrac{\partial\muhat}{\partial s}\right),
\end{equation}
while $\bLambda(s,t)$ enforces the constraint \eqref{eq:thatDef}. To solve for filament motion, we must equip these equations with appropriate boundary conditions (see \S\ref{sssec:ffmodel} and \S\ref{sssec:sfmodel}), as well as compute the fluid-filament interactions, both of which are done in the discrete setting.

\subsubsection{Numerical Discretisation}
The filament is discretised into $N$ segments of length $\Delta L=2.2 a$ such that $L=N\Delta L$. In all cases, the filaments have $N=20$ segments, yielding a filament aspect ratio $a/L=0.0227$. We discretise \eqref{eq:contBalance} using a second-order central difference method, and after multiplying the resulting equations by $\Delta L$, we obtain the discrete force and moment balances
\begin{equation}\label{eq:discBalance}
\begin{split}
    \boldsymbol{F}_n^C+\boldsymbol{F}_n^H+\boldsymbol{F}_n^A&=\boldsymbol{0},\\
    \boldsymbol{T}_n^E+\boldsymbol{T}_n^C+\boldsymbol{T}_n^H&=\boldsymbol{0},
\end{split}
\end{equation}
for each segment, $n$, where the superscripts $(\cdot)^E$, $(\cdot)^C$, $(\cdot)^H$, and $(\cdot)^A$ denote elastic, constraint, hydrodynamic and active forces and torques, respectively. We have
\begin{align}
    \boldsymbol{F}_n^C=&\bLambda_{n+1/2}-\bLambda_{n-1/2},\label{eq:FCLambda}\\
    \boldsymbol{T}_n^C=&\dfrac{\Delta L}{2}\that_n\times\left(\bLambda_{n+1/2}+\bLambda_{n-1/2}\right),\\
    \boldsymbol{T}_n^E=&\boldsymbol{M}_{n+1/2}-\boldsymbol{M}_{n-1/2}.
\end{align}
with half-indices representing the function values at the segment ends. The hydrodynamic force on segment $n$ is $\boldsymbol{F}_n^H=\boldsymbol{f}_n^H\Delta L$ and the torque is $\boldsymbol{T}_n^H=\boldsymbol{\tau}_n^H\Delta L$. The hydrodynamic forces and torques and the active force, $\boldsymbol{F}_n^A=\boldsymbol{f}_n^A\Delta L$, differ in the follower force and surface flow models, as we explain below.

\subsubsection{Follower Force Model}\label{sssec:ffmodel}
In the follower force model, filament motion is driven by the active forces, $\boldsymbol{F}^A_n$, on each $n$. We will consider two different distributions of follower force. For the tip-driven follower force model, the active force is applied only at the tip and is zero otherwise, namely, $\boldsymbol{F}^A_N=-F^A\that_N$ and $\boldsymbol{F}^A_n=\boldsymbol{0}$, $n\neq N$. In the distributed follower force model, we apply the same active force to each segment such that $\boldsymbol{F}^A_n=-F^A\that_n$ for all $n$.

Once the non-hydrodynamic forces are known, the resulting filament motion can be determined. Due to the low Reynolds number associated with biofilament motion, the hydrodynamic forces and torques on the filament segments will be linearly related to the translational velocity, $\boldsymbol{V}_n$, and angular velocity, $\boldsymbol{\Omega}_n$ on each segment $n$ through
\begin{equation}\label{eq:mobEq_ff}
    \begin{bmatrix}
        \boldsymbol{V}\\
        \boldsymbol{\Omega}
    \end{bmatrix}=\boldsymbol{\mathcal{M}}\,\begin{bmatrix}
        -\boldsymbol{F}^H\\
        -\boldsymbol{T}^H
    \end{bmatrix},
\end{equation}
where $\boldsymbol{V} = [\boldsymbol{V}_1^T,\boldsymbol{V}_2^T,\dots,\boldsymbol{V}_N^T]^T\in \mathbb{R}^{3N\times 1}$, $\boldsymbol{\Omega}=\left[\boldsymbol{\Omega}_1^T,\boldsymbol{\Omega}_2^T, \dots,\boldsymbol{\Omega}_N^T \right]^T\in\mathbb{R}^{3N\times1}$, and similarly, $\boldsymbol{F}^H,\,\boldsymbol{T}^H\in\mathbb{R}^{3N\times1}$ are the vectors containing all components of hydrodynamic force and torque, respectively, on all filament segments. Appearing in \eqref{eq:mobEq_ff} is the configuration dependent mobility matrix, $\boldsymbol{\mathcal{M}}\in\mathbb{R}^{6N\times6N}$ relating the forces and torques on the segments and the resulting motion. Following previous work \citep{schoeller2021,clarke2024}, we utilise the pairwise Rotne-Prager-Yamakawa (RPY) tensor \citep{wajnryb2013} with the hydrodynamic radius set to $a$ to provide the entries of the mobility matrix and hydrodynamic interactions between the segments.

Lastly, we must ensure that the resulting filament motion satisfies the boundary conditions at the clamped and free ends. In our discretisation, the clamped end condition is enforced by additional Lagrange multipliers to ensure $\bY_1(t)=\boldsymbol{0}$ and $\that_1(t)=\ehat_3$. At the free end, additional conditions are not needed as the follower force at the tip and moment-free condition are incorporated by setting $\boldsymbol{F}^A_N=-F^A\that_N$ with $\bLambda_{N+1/2}= 0$ and $\boldsymbol{M}_{N+1/2}=\boldsymbol{0}$, respectively.

\subsubsection{The Surface Flow Model}\label{sssec:sfmodel}
In the surface flow model where entrainment effects are included, we have $\boldsymbol{F}^A_n=\boldsymbol{0}$, for all $n$ and introduce activity through a mobility relation based on squirmer model \citep{lighthill1952,blake1971}. As a result of the surface flow, each segment $n$ attempts to move in the direction $-\that_{n}$ at speed $U = 2B_1/3$, where $B_1$ is the activity parameter that controls the surface flow strength. As a result, each active segment $n$ generates degenerate quadrupolar (potential dipolar) flow
\begin{equation}\label{eq:sqflow}
    \boldsymbol{u}_n(\boldsymbol{x})=\dfrac{1}{4\pi\eta r_{n}^3}\left(\idvec-3\,\rhat_{n}\rhat_{n}\right)\boldsymbol{H}_n
\end{equation}
where $\br_n=\boldsymbol{x}-\bY_n$, $r_n=\lVert r_n\rVert$, $\rhat_{n}=\br_n/r_n$ and $\boldsymbol{H}_n=4/3\pi a^3\,B_1\that_n$ is the quadrupole strength. Using this flow field, we can incorporate the effects of the surface flow into the segment mobility such that the resulting segment velocities are given by,
\begin{equation}\label{eq:mobEq_sf}
    \begin{bmatrix}
        \boldsymbol{V}\\
        \boldsymbol{\Omega}
    \end{bmatrix}=\boldsymbol{\mathcal{M}}\,\begin{bmatrix}
        -\boldsymbol{F}^H\\
        -\boldsymbol{T}^H
    \end{bmatrix}+\begin{bmatrix}
        \boldsymbol{\mathcal{S}}&\boldsymbol{0}\\\boldsymbol{0}&\boldsymbol{0}
    \end{bmatrix}\begin{bmatrix}
        \boldsymbol{H}\\
        \boldsymbol{0}
    \end{bmatrix},
\end{equation}
where $\boldsymbol{\mathcal{S}}\in\mathbb{R}^{3N\times3N}$ is the squirming matrix and $\boldsymbol{H}=[\boldsymbol{H}_1^T, \boldsymbol{H}_2^T, \dots, \boldsymbol{H}_N^T]^T \in \mathbb{R}^{3N\times1}$. The off-diagonal components of $\mathcal{S}$ are obtained by evaluating \eqref{eq:sqflow}, while the diagonal entries are of the form $-1/(2\pi\eta a^3)$ and account for the self-generated velocity. By changing $\boldsymbol{H}_n$, we can control the distribution of activity along the filament. For the tip-driven surface flow model, we take $\boldsymbol{H}_N=(4/3)\pi a^3B_1\that_N$ and $\boldsymbol{H}_n = 0$ for $n \neq N$. When actuation is uniform along the filament length in the distributed surface flow model, we have $\boldsymbol{H}_n=(4/3)\pi a^3B_1 \that_n$ for all $n$.

As in the follower force cases, we apply clamped-end boundary condition at the base by enforcing $\bY_1(t)=\boldsymbol{0}$ and $\that_1(t)=\ehat_3$. At the free end, we again have $\bLambda_{N+1/2}=\boldsymbol{0}$ and $\boldsymbol{M}_{N+1/2}=\boldsymbol{0}$.

\subsubsection{Differential-Algebraic System and Time-Stepping}
Once we have computed the segment translational and angular velocities, we can update their positions and orientations. For the orientations, we recognise that the local basis $\{\that_n,\muhat_n,\nuhat_n\}$ that is linked to filament deformation is related to the basis in the lab frame $\{\ehat_1,\ehat_2,\ehat_3\}$ through a rotation. Thus, we track the evolution of these vectors using unit-quaternions, $\boldsymbol{q}=[q_0,\bar{\boldsymbol{q}}]=[q_0,q_1,q_2,q_3]\in\mathbb{R}^4$ with $\lVert\boldsymbol{q}\rVert^2=q_0^2+q_1^2+q_2^2+q_3^2=1$. The quaternions map the standard basis to the local frame at each point along the filament via a matrix rotation $\boldsymbol{R}(\boldsymbol{q}_n)=\left[\that_n,\muhat_n,\nuhat_n\right]$, where
\begin{equation}
    \boldsymbol{R}(\boldsymbol{q})=\begin{bmatrix}
        1-2q_2^2-2q_3^2&2(q_1q_2-q_0q_3)&2(q_1q_3+q_0q_2)\\
        2(q_1q_2+q_0q_3)&1-2q_1^2-2q_3^2&2(q_2q_3-q_0q_1)\\
        2(q_1q_3-q_0q_2)&2(q_2q_3+q_0q_1)&1-2q_1^2-2q_2^2
    \end{bmatrix}.
\end{equation}

To update the positions and orientations of the segments, we must integrate the following nonlinear differential-algebraic system:
\begin{align}
    \dfrac{d\bY_n}{dt}=&\boldsymbol{V}_n,\label{eq:dYdt}\\
    \dfrac{d\boldsymbol{q}_n}{dt}=&\dfrac{1}{2}[0,\boldsymbol{\Omega}_n]\bullet\boldsymbol{q}_n,\label{eq:dqdt}\\
    \boldsymbol{0}=&\bY_{n+1}-\bY_n-\dfrac{\Delta L}{2}\left(\that_n+\that_{n+1}\right)\label{eq:Ytconstraint},
\end{align}
for each segment $n$, where $\boldsymbol{q}\bullet\boldsymbol{p}=[q_0,\bar{\boldsymbol{q}}]\bullet[p_0,\bar{\boldsymbol{p}}]=[q_0p_0-\bar{\boldsymbol{q}}\cdot\bar{\boldsymbol{p}},p_0\bar{\boldsymbol{q}}+q_0\bar{\boldsymbol{p}}+\bar{\boldsymbol{q}}\times\bar{\boldsymbol{p}}]$ is the quaternion product and the algebraic equations that must be satisfied are the discrete form of \eqref{eq:thatDef}. We employ a second-order geometric backward differentiation scheme to discretise the equations in time. We solve the resulting nonlinear system of equations using Broyden's method \citep{broyden1965} with a tolerance of $10^{-12}$. A detailed study of this numerical technique is presented in \cite{schoeller2021}.

To summarise, at each time step we obtain $\boldsymbol{M}$ from \eqref{eq:internalmoment}; $\boldsymbol{F}^H$ and $\boldsymbol{T}^H$ using \eqref{eq:discBalance}; $\boldsymbol{V}$ and $\boldsymbol{\Omega}$ from \eqref{eq:mobEq_ff} or \eqref{eq:mobEq_sf} depending on the model; $\bY$ from \eqref{eq:dYdt} and $\boldsymbol{q}$ from \eqref{eq:dqdt}. We treat $\bLambda$ as a Lagrange multiplier for this differential-algebraic system, so that at each time step, \eqref{eq:Ytconstraint} is satisfied. In our simulations, the typical time step size is $\Delta t=T/400$, where $T$ is the period of one cycle for periodic states, or based on the strongest frequency in non-periodic states.

\subsubsection{Non-Dimensional Parameters}
In the rest of this paper, we use the filament length $l^*=L$ as the length scale, relaxation time $t^*={\eta\,L^4}/{K_B}$ as the time scale, and $F^*={K_B}/{L^2}$ as the force scale. Accordingly, we introduce
\begin{equation}
    f=\dfrac{F^AL^2}{K_B},\qquad b_1=\dfrac{\eta L^3B_1}{K_B},
\end{equation}
as the non-dimensional actuation parameters for follower force models and surface flow models, respectively.

\subsection{Dynamical Systems Analysis}
In this section, we briefly describe the computational tools we employ to perform a dynamical systems analysis of the filament models. In order to use these tools, we must first introduce an appropriate state vector. Unfortunately, the unit-quaternions are not suitable since the sum of two unit-quaternions is not itself a unit-quaternion. In our case, it is therefore convenient to instead employ the effective Lie algebra element \citep{clarke2024}, $\bU_n\in\mathbb{R}^3$, for each segment $n$ to describe the rotation of the frame $\left\{\that_n,\muhat_n,\nuhat_n\right\}$ relative to a state where the filament is straight and upright such that $\that_n=\ehat_3$, $\muhat_n=\ehat_1$, and $\nuhat_n=\ehat_2$ for all $n$. In simple terms, an effective Lie algebra element is a 3D generalisation of the tangent angle formulation that can be employed in 2D. Since the quaternion corresponding to the upright state is
\begin{equation}
    \boldsymbol{q}_\text{eq}=\begin{bmatrix}
        \dfrac{1}{\sqrt{2}},0,-\dfrac{1}{\sqrt{2}},0
    \end{bmatrix},\qquad\text{with}\qquad\boldsymbol{q}^*_\text{eq}=\begin{bmatrix}
        \dfrac{1}{\sqrt{2}},0,\dfrac{1}{\sqrt{2}},0
    \end{bmatrix}.
\end{equation}
and the rotation of $n$ relative to this state is given by the quaternion product $\boldsymbol{p}=\boldsymbol{q}_n\bullet\boldsymbol{q}^*_\text{eq}=\begin{bmatrix}
    p_0,\bar{\boldsymbol{p}}
\end{bmatrix}$, the effective Lie algebra element associated with a quaternion $\boldsymbol{q}_n$ is
\begin{equation}
    \bU_n = 2\arccos{\left(p_0\right)}\hat{\boldsymbol{p}},
\end{equation}
where $\hat{\boldsymbol{p}}={\bar{\boldsymbol{p}}}/{\lVert\bar{\boldsymbol{p}}\rVert}$. The state vector is then given by $\bU = [\bU_1^T,\bU_2^T,\dots,\bU_N^T]^T\in\mathbb{R}^{3N\times1}$ with $\bU_1 = 0$ to incorporate the clamped end condition. We refer the reader to Appendix~\ref{app:effLieAlgEl} for more about properties of the effective Lie algebra element.

\subsubsection{Jacobian-Free Newton-Krylov (JFNK) Method}\label{sssec:jfnk}
The filament models display several time-periodic solutions after the steady, upright state becomes unstable. To find these time-periodic solutions, we first shift our thinking and consider the filament simulation as a means of computing the flow $\boldsymbol{\phi}^t(\bU)$ that maps solutions $\bU(0)$ to $\bU(t)$. We then seek effective Lie algebra elements that satisfy
\begin{equation}\label{eq:jfnk1}
    \boldsymbol{F}(\bU,T)=\boldsymbol{\phi}^T(\bU)-\bU=\boldsymbol{0}
\end{equation}
for some non-zero, but unknown period, $T$. We obtain solutions to \eqref{eq:jfnk1} using a Newton-Krylov method with a GMRES-hook step, described in \cite{viswanath2007,viswanath2009}. Our implementation closely follows that of \cite{willis2019}. We continue the iterations until $\lVert \boldsymbol{F}(\bU,T)\rVert<10^{-10}$, with $\lVert\bcdot\rVert$ as the 2-norm operator.

\subsubsection{Linear Stability Analysis}
Along with finding time-periodic solutions, we would like assess their stability, as well as the stability of the steady, upright state. The time evolution of the state vector can be expressed as
\begin{equation}\label{eq:lsa1}
    \dfrac{d \bU}{d t}=\boldsymbol{f}(\bU;t).
\end{equation}

In linear stability analysis, we are interested in the evolution of small perturbations $\delta\bU$ around a (steady or time-periodic) base state $\bU_0$. Growing perturbations indicate unstable base states, whereas dampened perturbations indicate linearly stable base states. We linearise \eqref{eq:lsa1} around $\bU_0$ and obtain the linearised dynamics for the perturbation as
\begin{equation}\label{eq:lsa2}
    \dfrac{d\delta\bU}{d t}=\boldsymbol{A}\delta\bU,
\end{equation}
where $\boldsymbol{A}=\left.\dfrac{\partial\boldsymbol{f}}{\partial\bU}\right|_{\bU_0}$.

Equation \eqref{eq:lsa2} is linear and has the principal fundamental matrix solution $\boldsymbol{\Phi}(t;\bU_0(t))$ such that the general solution is $\delta\bU(t)=\boldsymbol{\Phi}(t;\bU_0(t))\,\delta\bU_0$. For a steady base state with $d\bU_0/dt=\boldsymbol{0}$,
\begin{equation}\label{eq:lsa3}
    \boldsymbol{\Phi}(t;\bU_0(t))=e^{t\boldsymbol{A}}.
\end{equation}

If $\bU_0$ is instead time periodic such that $\bU_0(t)=\bU_0(t+T)$, Floquet's theorem states that there is a Floquet normal form of the fundamental solution matrix,
\begin{equation}\label{eq:lsa4}
    \boldsymbol{\Phi}(t;\bU_0)=\boldsymbol{P}(t)e^{t\boldsymbol{B}},
\end{equation}
with a $T$-periodic vector matrix $\boldsymbol{P}(t)$ and a matrix operator $\boldsymbol{B}$ \citep{perko2008}. Notice that under the action of linearised Poincar{\'e} maps $\bU(t)\mapsto\bU(t+T)$, a periodic state is steady and the matrix $\boldsymbol{B}$ is analogous to the matrix $\boldsymbol{A}$ of the steady case.

The stability of a base state depends on the eigenvalues $\lambda_i$ of $\boldsymbol{A}$ or $\boldsymbol{B}$. A solution is linearly stable if all of its eigenvalues have nonpositive real parts. We calculate the eigenvalues $\mu_i$ of $\boldsymbol{\Phi}(T;\bU_0)$ where $T$ is the period for periodic states and a sufficiently small time for steady base states. We compute the eigenvalues of $\boldsymbol{\Phi}(T;\bU_0)$ using Arnoldi iteration, which does not require explicit computation of $\boldsymbol{\Phi}(T;\bU_0)$ but only its action on a set of vectors. At each Arnoldi iteration, an upper Hessenberg matrix is calculated, whose eigenvalues approximate that of $\boldsymbol{\Phi}(T;\bU_0)$. We record the first 13 eigenvalues as a vector at each Arnoldi iteration and continue until the norm of the difference between consecutive iterations is lower than $10^{-8}$.

The eigenvalues $\lambda_i$ are related to the eigenvalues of $\boldsymbol{\Phi}(T;\bU_0)$ via
\begin{equation}\label{eq:lsa5}
    \lambda_i=\dfrac{1}{T}\log{(\mu_i)}.
\end{equation}

\subsubsection{Spectral Proper Orthogonal Decomposition (SPOD)}\label{sssec:spod}
Along with the upright steady-state and time-periodic solutions, the filament models can also admit more complex solutions, such as quasi-periodic states. To study these states, we extract their dominant motions at each frequency $\omega$ using spectral proper orthogonal decomposition (SPOD). We utilise the implementation developed in \cite{towne2018}, and summarise the procedure in our context below.

Given effective Lie algebra elements $\bU(s,t)$, $\boldsymbol{V}(s,t)$ and the inner product $\langle\bU,\boldsymbol{V}\rangle_{s,t}=\int\limits_{-\infty}^\infty\int\limits_0^L\boldsymbol{V}^*(s,t)\,\bU(s,t)\,dsdt$, with the superscript $(\cdot)^*$ denoting the Hermitian transpose, SPOD modes are obtained by solving the following optimisation problem:
\begin{equation}\label{eq:spod2}
    \max_\phi\dfrac{E\{\lvert\langle\bU(s,t),\boldsymbol{\phi}(s,t)\rangle_{s,t}\rvert^2\}}{\langle\boldsymbol{\phi}(s,t),\boldsymbol{\phi}(s,t)\rangle_{s,t}},
\end{equation}
where $E\{\cdot\}$ denotes ensemble average, for some function $\boldsymbol{\phi}(s,t)$ that approximates $\bU(s,t)$ on average. With a Fourier transform $\boldsymbol{\phi}(s,t)=\int_{-\infty}^\infty \boldsymbol{\psi}(s,\omega)\,e^{i2\pi\omega t} d\omega$, the optimisation problem in \eqref{eq:spod2} can be formulated at each frequency $\omega$. The solution to the optimisation problem is obtained by solving the following eigenvalue problem:
\begin{equation}\label{eq:spod3}
    \int\limits_0^L\boldsymbol{S}(s,s^\prime,\omega)
    \,\boldsymbol{\psi}(s^\prime,\omega)\,ds^\prime=\lambda(\omega)\,\boldsymbol{\psi}(s,\omega),
\end{equation}
where $\boldsymbol{S}(s,s^\prime,\omega)$ is the cross-spectral density tensor, defined as the Fourier transform of the two-point space-time correlation tensor. For further details, the reader may refer to \cite{towne2018}.

Given time series data of the effective Lie algebra element, we block the data into multiple realisations using Welch's method and take the Fourier transform with a Hamming window to reduce spectral leakage due to non-periodicity of each block. By ensemble averaging these Fourier realisations, the cross-spectral density matrix is obtained for each frequency, and the SPOD modes are calculated by performing an eigenvalue decomposition of the covariance operator in the frequency domain. The SPOD analysis based on these methods are performed using the software from \cite{towne2018} available at \url{https://github.com/SpectralPOD/spod_matlab}.

\subsubsection{Bisection Method}
As we discuss in \S\ref{sec:ResultsandDiscussions}, we find a range of actuation for which filament dynamics exhibits bistability. Here, the two periodic states of whirling and beating are connected through an unstable quasiperiodic state, QP1. To compute QP1, we use the bisection algorithm from \cite{skufca2006} and \cite{schneider2009}.

Given two initial conditions $\bU_\mathrm{w}$ and $\bU_\mathrm{b}$ that converge with time to whirling and beating, respectively, we compute the bisected initial condition as
\begin{equation}\label{eq:biseq1}
    \bU_\mathrm{QP}=\dfrac{\bU_\mathrm{w}+\bU_\mathrm{b}}{2}.
\end{equation}
With this initial condition, we evolve the system to long times and check if the filament dynamics are attracted to either of the periodic states. Once the solution reaches (criteria provided in Appendix~\ref{app:bisec}) either of the states, we update $\bU_\mathrm{w}$ or $\bU_\mathrm{b}$ accordingly and perform a bisection again according to \eqref{eq:biseq1}. We repeat this process until $\lVert\bU_\mathrm{w}-\bU_\mathrm{b}\rVert<10^{-8}$.

\section{Results and Discussions}\label{sec:ResultsandDiscussions}
We investigate the state space of the four different active filament models as we vary the actuation. Follower force models use the dimensionless follower force magnitude, $f$, as the actuation parameter, while surface flow models use the dimensionless surface flow magnitude, $b_1$. These two parameters are directly related. A segment with follower force magnitude $f$ has active force $F^A=fK_B/L^2$, while a surface flow segment with $b_1$ experiences drag $F_d=4\pi ab_1K_B/L^3$. Equating these two forces yields that the follower force and the surface flow are related through $b_1\approx3.5f$.

In exploring filament dynamics, we divide our presentation of the results into two actuation ranges: low-to-moderate and high. Primary and secondary filament instabilities occur within the low-moderate actuation regime, where previous studies have revealed similar filament dynamics \citep{laskar2017,decanio2017,ling2018,clarke2024}. At high actuation, these states are unstable and we observe greater differences in the dynamics given by the models.

\subsection{Dynamic states at Low-to-Moderate Actuation}\label{ssec:stateslow}
\begin{figure}
    \centering
    \includegraphics[width=0.99\textwidth]{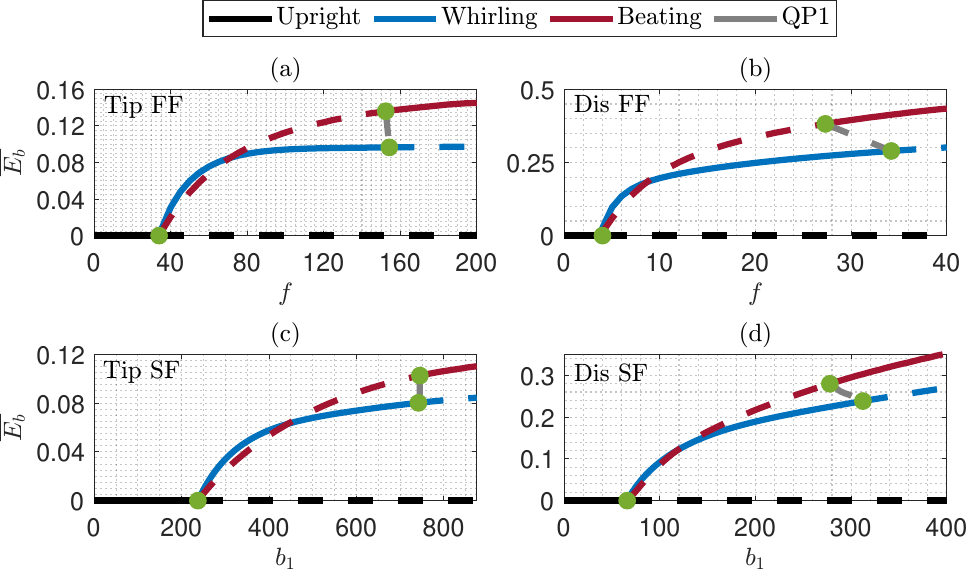}
    \caption{Bifurcation diagrams up to moderate actuation for (a) tip-driven follower force, (b) distributed follower force, (c) tip-driven surface flow, and (d) distributed surface flow models. Stable states are marked in solid lines, whereas unstable ones are in dashed lines. We observe similar bifurcation patterns and states in low to moderate actuation levels. We plot the non-trivial filament states in Figure~\ref{fig:statessfdis}.}
    \label{fig:bifdiag}
\end{figure}

We begin by classifying the dynamic states up to moderate actuation and expand on previous results by directly comparing the different models. As stated earlier, we consider two actuation models: follower force (FF) and surface flow (SF), and two actuation profiles: concentrated at the tip (Tip) and uniformly distributed along the length of the filament (Dis). In Figure~\ref{fig:bifdiag} we present the bifurcation diagram for each model up to moderate actuation, which shows the possible filament states and transitions between these states. We indicate linearly stable states with solid lines and unstable states with dashed lines. We use the dimensionless mean bending energy,
\begin{equation}
    \overline{E}_b=\dfrac{1}{T}\int\limits_0^T\int\limits_0^L\dfrac{1}{2}\dfrac{\partial^2\bY(s,t)}{\partial s^2}\bcdot\dfrac{\partial^2\bY(s,t)}{\partial s^2}dsdt,
\end{equation}
to distinguish different states. By comparing Figure~\ref{fig:bifdiag}(a,c) with (b,d), we see that the bending energy is overall lower for the tip-driven cases than for the distributed cases. This implies that models with a distributed actuation profile produce filament states with higher curvature.
\begin{figure}
    \centering
    \includegraphics[width=0.99\textwidth]{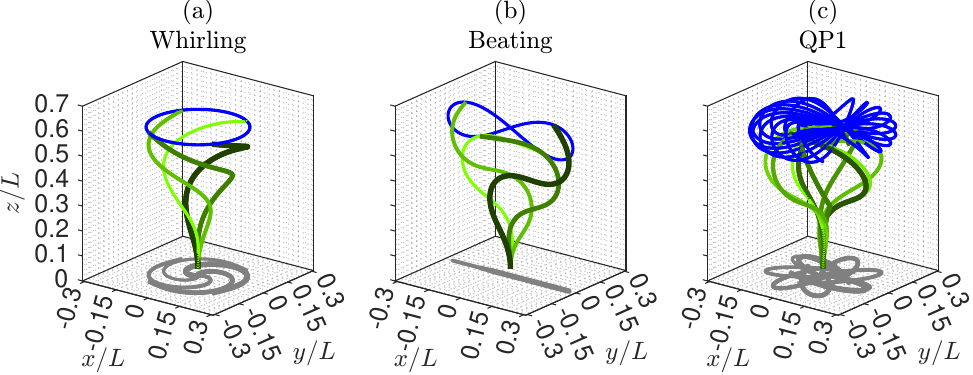}
    \caption{Non-trivial states observed in all active filament models up to moderate actuation: (a) whirling, (b) beating, (c) QP1. All states are given at $b_1=284$ for the distributed surface flow model (Dis SF). The filament's colour gets darker as time passes. The filament's shadow is given in gray and its tip pathline is given in blue. We provide a video description of the states in movie 1.}
    \label{fig:statessfdis}
\end{figure}

Apart from small quantitative differences, the filament models exhibit similar states and dynamics in the low-moderate actuation regime. In all active filament models, we observe four common states: upright, whirling, beating, and QP1. We plot these states (except upright) in Figure~\ref{fig:statessfdis}. The increasing direction of the time is indicated by darker filament colours. The pathline of the filament tip is a closed curve for periodic states, whereas for QP1 it lies on a torus. To aid the viewer, we also plot the shadow of the filament at its base. Upright is a steady state (fixed point); whirling and beating are time-periodic; and QP1 is a quasiperiodic state.

From the bifurcation diagram, Figure~\ref{fig:bifdiag}, we see that regardless of the type and distribution of actuation, we observe a double Hopf bifurcation when the upright state becomes unstable, creating whirling and beating (see \S\ref{ssec:piuf} for further details). Later, whirling undergoes a Hopf bifurcation and beating undergoes a pitchfork bifurcation. QP1 connects to whirling at one end and beating at the other (see \S\ref{ssec:wbqp1}) through these bifurcations.

\subsection{Primary Instability of the Upright Filament}\label{ssec:piuf}
\begin{figure}
    \centering
    \includegraphics[width=0.90\textwidth]{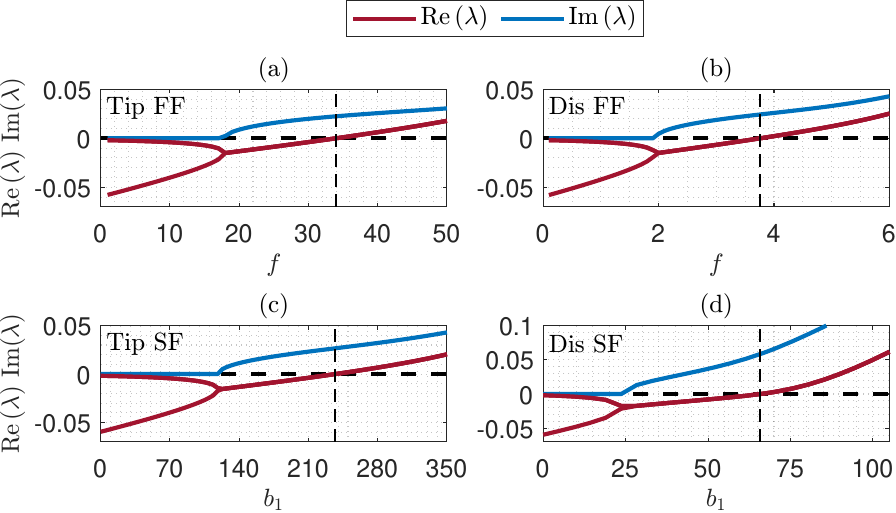}
    \caption{Leading two eigenvalues of the upright state for (a) tip-driven follower force, (b) distributed follower force, (c) tip-driven surface flow, and (d) distributed surface flow models. Real and imaginary parts of the eigenvalues are shown in red and blue, respectively. The double Hopf bifurcation is marked with a vertical dashed line.}
    \label{fig:eigvals_upright}
\end{figure}

The upright filament is the trivial state and corresponds to a straight filament aligned with $\ehat_3$. Actuation produces internal stresses and like beams under compression, when these stresses exceed a threshold, the filament buckles. While the critical actuation value is model dependent, this instability is observed in each model. We compute the eigenvalues of the linearised operator around the upright state and in Figure~\ref{fig:eigvals_upright} plot the first two eigenvalues for each model as functions of the actuation parameter. The real and imaginary parts of the eigenvalues are coloured red and blue, respectively. The vertical dashed line indicates when the real part of the leading eigenvalue is zero. The leading eigenvalues and their dependence on the actuation parameter are qualitatively similar in all cases. We note, however, that the critical value of actuation is higher in the surface flow models. Consistent with \cite{decanio2017}, this indicates that the hydrodynamic entrainment provides a mechanism that delays the bifurcation.

In all cases, as the actuation increases, we observe that identical sets of two real eigenvalue branches (only one set is shown in Figure~\ref{fig:eigvals_upright}) merge. This point marks the transition of the upright filament from a stable node to a stable focus, i.e., the two sets of two real eigenvalues transition to two sets of a complex conjugate eigenvalue pair. Due to the symmetry of the upright state and no preferred directionality in the $xy$-plane, the leading eigenvalues are identical complex-conjugate pairs with one eigenmode corresponding to bending in the $xz$-plane, and the other to bending the $yz$-plane.

These two sets of a complex conjugate eigenmodes eventually become unstable at the same actuation value. As described in \cite{clarke2024}, when this occurs, two periodic states, whirling and beating, emerge through a double Hopf bifurcation with whirling being stable and beating unstable. In all models, whirling involves a filament with time-independent curvature rotating about the $z$-axis at a constant speed (Figure~\ref{fig:statessfdis}(a)), while beating is a planar state where the tip traces an infinity symbol-shaped path (Figure~\ref{fig:statessfdis}(b)). These states have been observed in the tip-driven follower force model \citep{chelakkot2014,decanio2017,clarke2024} and the tip-driven active colloid model \citep{laskar2013,laskar2017}.

As stated above, due to the rotational symmetry of the upright state, there are two decoupled orthonormal eigenmodes at the instability threshold, which individually describe planar beating oscillations. Without loss of generality, we can define these two orthonormal eigenmodes using the effective Lie algebra element: one for the $xz$-plane, $\boldsymbol{\nu}_x(s)=[0,v(s),0]^T$, and one for the $yz$-plane, $\boldsymbol{\nu}_y(s)=[v(s),0,0]^T$, where $v(s)$ describes the filament shape. For surface flow models, periodic solutions close to the bifurcation take the form
\begin{equation}
    \bU(s,t)\approx\left(A_x\boldsymbol{\nu}_x(s)+A_y\boldsymbol{\nu}_y(s)\right)\sqrt{b_1-b_1^*}\,e^{i\omega^*t}+cc,
\end{equation}
where $b_1^*$ and $\omega^*$ are the surface flow magnitude and the frequency at the bifurcation, respectively. The amplitudes $A_x, A_y\in\mathbb{C}$ are determined from the initial value $\bU(s,0)$ of the periodic state. The form of the solution is the same for the follower force model with $b_1$ replaced by $f$ (see \cite{clarke2024}).

\begin{figure}
    \centering
    \includegraphics[width=0.95\textwidth]{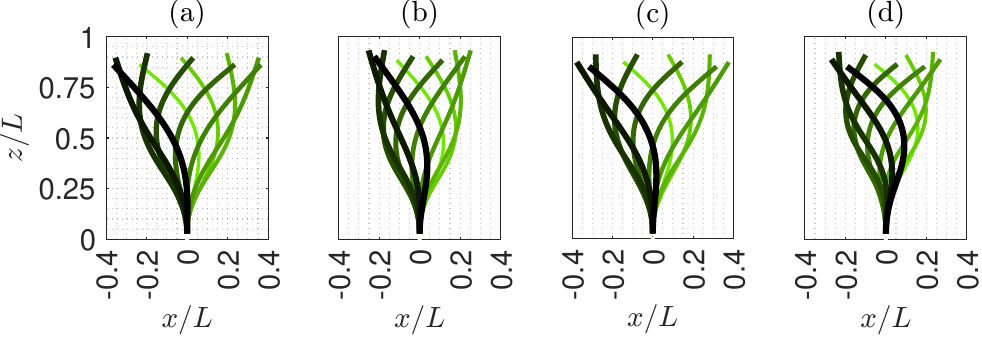}
    \caption{Leading eigenmodes at the double Hopf bifurcation for (a) tip-driven follower force, (b) distributed follower force, (c) tip-driven surface flow, and (d) distributed surface flow models. The plots show the eigenmode on $xz$-plane at different phases. Linear combination of the mode with its counterpart on the $yz$-plane creates beating state, while a linear combination with a $\pi/2$ phase-lag creates whirling state, as depicted in Figure~\ref{fig:statessfdis}.}
    \label{fig:hhEigMode}
\end{figure}

For the motion to remain planar, the eigenmodes must remain in phase, so $\operatorname{Im}\left(\langle A_x,A_y\rangle\right)=0$ for beating. For whirling, $A_y = iA_x$, meaning one mode lags behind the other by a $\pi/2$ phase. These results are derived using the connection between filament states and effective Lie algebra elements in Appendix~\ref{app:effLieAlgEl}.

In Figure~\ref{fig:hhEigMode}, we plot the leading eigenmode at the bifurcation in the $xz$-plane at different phases. The curved filament shape in each model is associated with its buckling, however, there are some qualitative differences between the modes for the different models.

We investigate further the buckling instability by considering the force balance in the upright state. Linear stability analysis using resistive force theory \citep{decanio2017,ling2018,schnitzer2025} reveals that the internal stress distribution appears in the linear operator, and hence contributes to the eigenvalues of the primary instability. Motivated by this, we determine the internal stress for each model, with the aim of explaining the differences in the mode shapes in Figure~\ref{fig:hhEigMode}.

In follower force models, the hydrodynamic force and torque are zero due to the absence of induced surface flow and the filament being at rest. Thus, the internal force results solely from balancing the prescribed active force, given by $\bLambda(s)=\bLambda(L)+\int_s^L\boldsymbol{f}^A(s')ds'$ with the appropriate boundary condition at $\bLambda(L)$. For the tip-driven case, we take $\boldsymbol{f}^A(s)=\boldsymbol{0}$ and $\bLambda(L)=-F^A\,\ehat_3$ where $F^A$ is the prescribed dimensional follower force magnitude. We note that this is equivalent to our numerical framework where we instead activity is incorporated through the body force rather than the boundary condition, see \cite{ling2018}.  Consequently, the internal force distribution is $\bLambda(s)=-F^A\,\ehat_3$. In the distributed case, there is a uniform force distribution $\boldsymbol{f}^A(s)=-F^A/\Delta L\,\ehat_3$. Hence, $\bLambda(s)=-F^A(L-s)/\Delta L\,\ehat_3$.

In the surface flow model, however, the hydrodynamics play a role as the internal stress is obtained by solving a resistance problem. For the surface flow model, the filament motion is given by
\begin{equation}
    \begin{bmatrix}
        \boldsymbol{V}-\boldsymbol{V}^\mathrm{sf}\\
        \boldsymbol{\Omega}
    \end{bmatrix}=\boldsymbol{\mathcal{M}}\begin{bmatrix}
        -\boldsymbol{F}^H\\
        -\boldsymbol{T}^H
    \end{bmatrix},
\end{equation}
where $\boldsymbol{V}^\mathrm{sf}=\mathcal{S}\boldsymbol{H}$ is the surface velocity. In the upright state, $\boldsymbol{V}=\boldsymbol{0}$ and $\boldsymbol{\Omega}=\boldsymbol{0}$, yet non-zero hydrodynamic forces arise due to surface flow activity. As a result, internal stresses are given by the resistance problem,
\begin{equation}\label{eq:invmobprob}
    \begin{bmatrix}
        \boldsymbol{F}^C\\\boldsymbol{T}^C
    \end{bmatrix}=-\boldsymbol{\mathcal{M}}^{-1}\begin{bmatrix}
        \boldsymbol{V}^\mathrm{sf}\\\boldsymbol{0}
    \end{bmatrix}.
\end{equation}
After solving this linear system for the tip-driven and distributed surface velocities, the internal stress can be determined by relating $\boldsymbol{F}^C$ to $\bLambda$ using \eqref{eq:FCLambda}.

\begin{figure}
    \centering
    \includegraphics[width=0.70\textwidth]{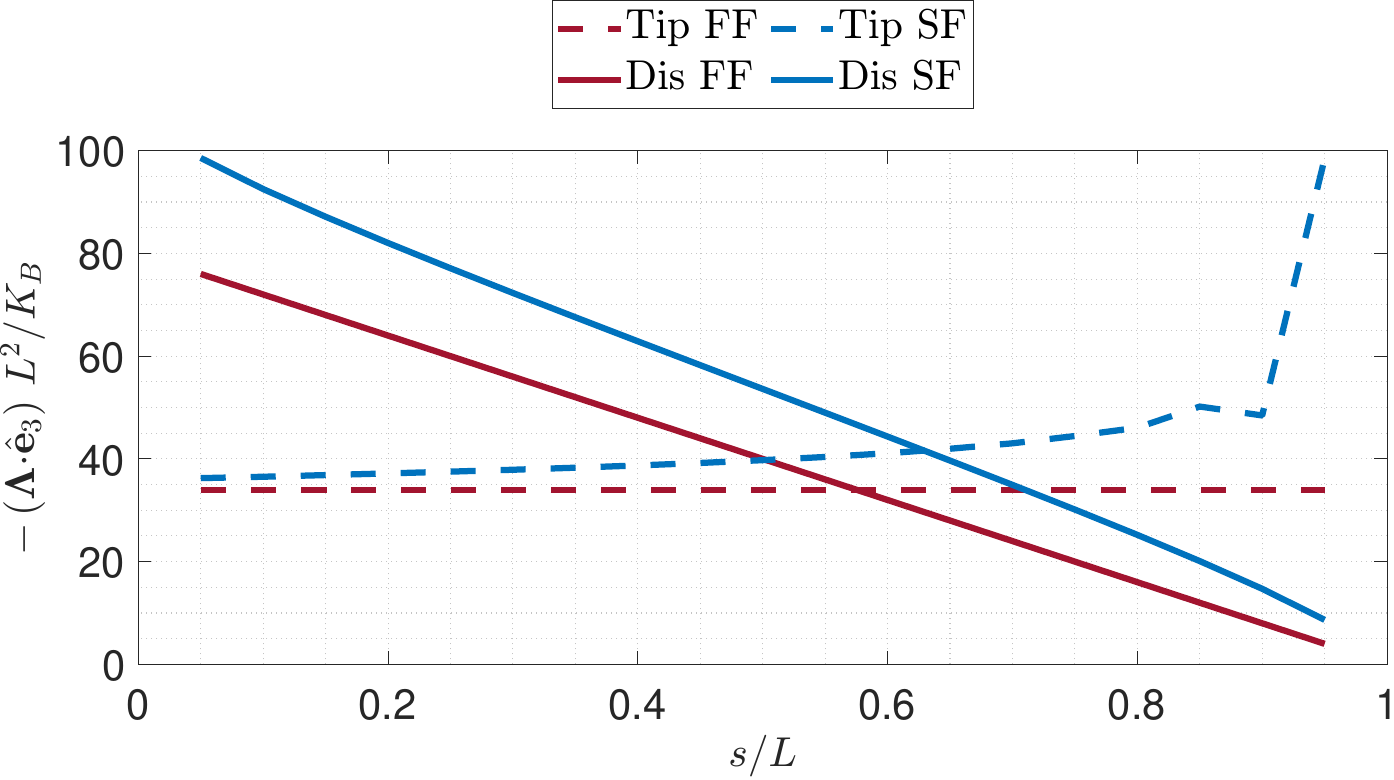}
    \caption{Force distribution along the filament for different active filament models at the double Hopf bifurcation. Follower force models are plotted in red, while surface flow models are plotted in blue. Tip-driven models are given in dashed lines, whereas distributed models are plotted in solid lines.}
    \label{fig:hhForceDist}
\end{figure}

In Figure~\ref{fig:hhForceDist}, we plot the internal stress along the filament at buckling for the different models. For the distributed models, the stress is accumulated at the base and decays monotonically as one moves toward the free-end. For the tip-driven follower force model, the stress in constant along the length. For the tip-driven surface flow case, however, after first increasing only gradually with arclength, the stress exhibits a sudden peak at the free end. This peaking in the stress is reminiscent of the distribution of hydrodynamic force on slender bodies and high aspect ratio prolate spheroids translating in fluid \citep{kim2013}. Here, it arises due to the resistance problem that yields the internal stress.

As a result of the differences in these stress distributions, the eigenmodes for tip-driven models have a wider tip span (see Figure~\ref{fig:hhEigMode}(a,c)), whereas distributed models show narrower tips and wider bases (see Figure~\ref{fig:hhEigMode}(b,d)). Figure~\ref{fig:hhForceDist} also shows that the filaments buckle when internal stresses exceed a threshold, which is generally higher for surface flow models. The surface flow, in opposing direction of compression, allows the filament to support larger loads, thus delaying the bifurcation.

\subsection{Secondary Instabilities and Quasiperiodic Transition}\label{ssec:wbqp1}
The double Hopf bifurcation creates two limit cycles: whirling and beating. In this section, we examine the instabilities associated with these periodic states. Near the double Hopf bifurcation, whirling is stable, while beating is unstable. However, as shown in Figure~\ref{fig:bifdiag}, beating becomes stable and whirling unstable at moderate actuation. We focus on understanding the changes in the stability of these periodic states and the origin of this transition.

Using the JFNK algorithm described in \S\ref{sssec:jfnk}, we track each state as the actuation increases. Increased actuation leads to higher internal stresses, increasing filament deformation and its bending energy, as shown in Figure~\ref{fig:bifdiag}(a) for the tip-driven follower force model. For whirling, the radius of the circular path of each point along the filament increases, while for beating, the filament sweeps out a greater area. We also observe that periodic states tend to have higher frequencies as the actuation increases.

\begin{figure}
    \centering
    \includegraphics[width=0.90\textwidth]{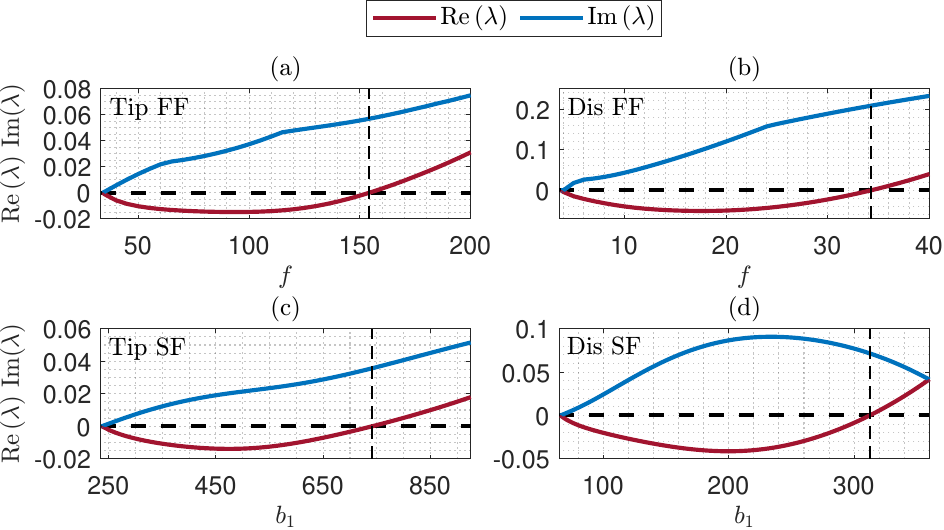}
    \caption{Leading eigenvalue of the whirling state for (a) tip-driven follower force, (b) distributed follower force, (c) tip-driven surface flow, and (d) distributed surface flow models. Real and imaginary parts of the leading eigenvalues are shown in red and blue, respectively. The Hopf bifurcation is marked with a vertical dashed line.}
    \label{fig:eigvals_whirling}
\end{figure}

\begin{figure}
    \centering
    \includegraphics[width=0.95\textwidth]{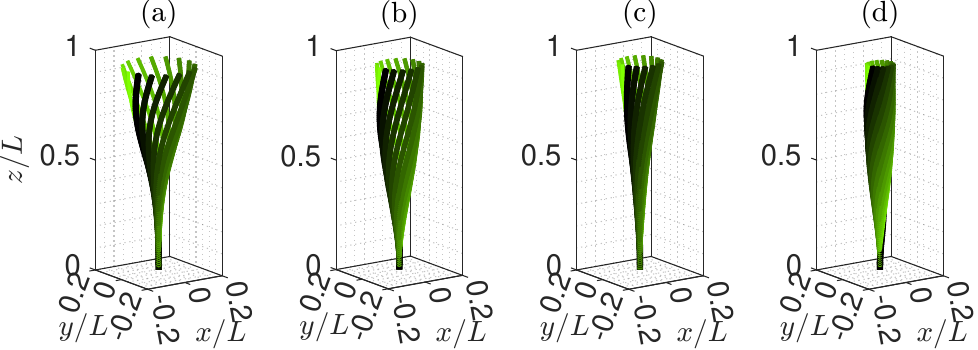}
    \caption{Leading eigenmodes at the Hopf bifurcation of whirling for (a) tip-driven follower force, (b) distributed follower force, (c) tip-driven surface flow, and (d) distributed surface flow models. Different filament colours correspond to different phases of the mode.}
    \label{fig:hbEigMode}
\end{figure}

We perform a linear stability (Floquet) analysis for each state along these branches. The leading eigenvalues from the linear stability analysis of the whirling state are shown in Figure~\ref{fig:eigvals_whirling}. The leading eigenvalue is complex, meaning that whirling becomes unstable through a Hopf bifurcation. We plot the leading eigenvector of the monodromy matrix at different phases for each filament model in Figure~\ref{fig:hbEigMode}. These eigenvectors represent the instability dynamics, with the time-periodicity of whirling removed. Comparing the mode shapes for the tip-driven and distributed models (Figures~\ref{fig:hbEigMode}(a,b) and (c,d)), we observe that the unstable mode shape narrows toward the filament tip for distributed models, while it is widest at the tip for tip-driven models.

\begin{figure}
    \centering
    \includegraphics[width=0.80\textwidth]{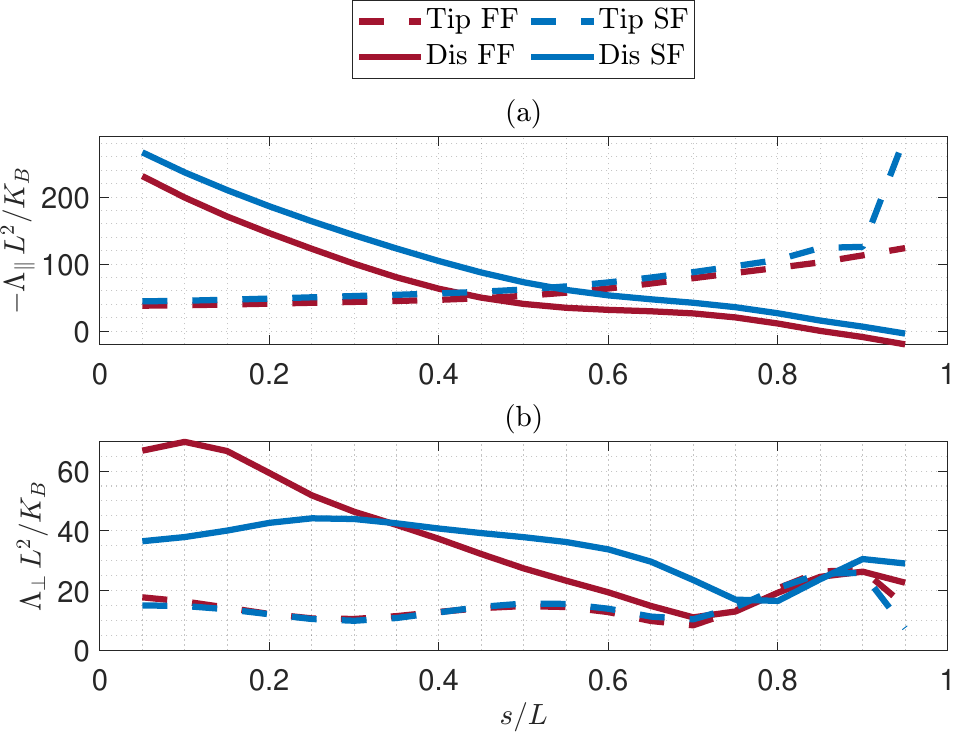}
    \caption{(a) Tension ($\Lambda_\parallel=\bLambda\bcdot\that$) and (b) shear ($\Lambda_\perp=\lVert\bLambda-\Lambda_\parallel\that\rVert$) distribution along the length of a filament for different active filament models at the Hopf bifurcation. Follower force models are plotted in red, while surface flow models are plotted in blue. Tip-driven models are given in dashed lines, whereas distributed models are plotted in solid lines.}
    \label{fig:h1ForceDist}
\end{figure}

Since whirling can be viewed as a steady state in a rotating frame of reference, compression and shear forces are time-invariant in that frame. Figure~\ref{fig:h1ForceDist} shows the force distribution for each active filament model. Compressive forces are transferred along the filament and balanced by the constraint force at the base. Shear is balanced by elastic forces that arise due to filament curvature, along with the viscous forces as a result of filament motion. As for the steady state profile, we observe that distributed models have an almost linearly decaying tangential stress distribution (see Figure~\ref{fig:h1ForceDist}(a)). Due to its suspected presence in the linearised operator, this distribution of tangential stress produces to a bottom-heavy mode shape, as seen in Figure~\ref{fig:hbEigMode}(d) for the distributed surface flow model. In contrast, the tip-driven models exhibit an increase in the tangential stress with arclength. Again, the surface flow model exhibits an additional peak at the free-end. This form of the distribution results in a mode shape with a wide tip (Figure~\ref{fig:hbEigMode}(c)). Comparing Figure~\ref{fig:h1ForceDist}(b) with (a), we observe that shear is much weaker than compression.

\begin{figure}
    \centering
    \includegraphics[width=0.90\textwidth]{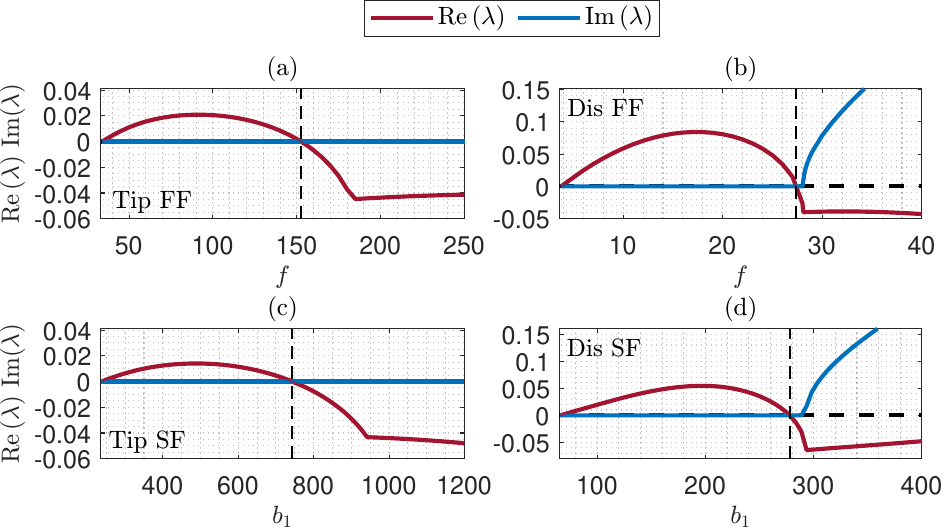}
    \caption{Leading eigenvalue of the beating state for (a) tip-driven follower force, (b) distributed follower force, (c) tip-driven surface flow, and (d) distributed surface flow models. Real and imaginary parts of the leading eigenvalues are shown in red and blue, respectively. The pitchfork bifurcation is marked with a vertical dashed line.}
    \label{fig:eigvals_beating}
\end{figure}

\begin{figure}
    \centering
    \includegraphics[width=0.95\textwidth]{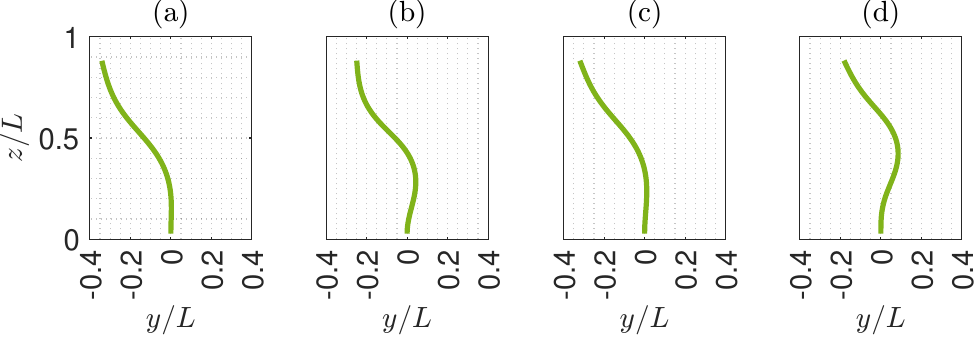}
    \caption{Leading eigenmodes at the pitchfork bifurcation for (a) tip-driven follower force, (b) distributed follower force, (c) tip-driven surface flow, and (d) distributed surface flow models. The base beating state for each model is at the same phase (when the tip of the filament is at its leftmost position) to remove phase dependency of the Floquet modes. The instability mechanism is out-of-plane tipping, as the most unstable mode is a planar filament state lying orthogonal to the beating plane.}
    \label{fig:pbEigMode}
\end{figure}

We also assess the stability of planar beating which appears in all models as an unstable solution after the filament buckles. Figure~\ref{fig:eigvals_beating} shows the leading eigenvalues from the stability analysis for the beating state. In all cases, the leading eigenvalue is purely real, so beating becomes stable through a pitchfork bifurcation. For our stability calculations, we use base states that beat in the $xz$-plane. Additionally, as the resulting eigenmodes will be dependent on the phase of the base state, we set $t=0$ in our stability calculations to correspond to the instance when the filament tip attains its maximum amplitude in the $-x$-direction. The unstable modes at the bifurcation for each model are shown in Figure~\ref{fig:pbEigMode}. We see that the eigenmodes are planar, but in the $yz$-plane, orthogonal to the base state. This indicates that as actuation decreases, planar beating goes unstable through out-of-plane tipping.

These two bifurcations transition filament dynamics from a range where only whirling is stable to one where only beating is stable. Bridging these two bifurcations is the quasiperiodic state, QP1, see Figure \ref{fig:bifdiag}.  In this study, we find that while QP1 is a generic feature of active filament models, its stability appears to be sensitive to the details of the model. In the tip-driven surface flow model, whirling and beating undergo supercritical bifurcations, with a stable QP1 emerging from the Hopf bifurcation and annihilating at the pitchfork bifurcation. In contrast, all other models feature subcritical bifurcations, leading to bistability between whirling and beating and an unstable QP1. In these cases, we obtain the unstable QP1 state using the bisection algorithm. From Figure~\ref{fig:bifdiag}, we observe that distributed models have a wider range of bistability.  In \cite{clarke2024}, the tip-driven follower force filament anchored to a no-slip surface was instead found to exhibit a stable QP1 state, indicating that this change in hydrodynamics is sufficient to change the stability of QP1.  In addition, it has been shown that even in an unbounded fluid increasing the filament aspect ratio by adding more segments in the discrete filament model is also able to stabilise QP1 \citep{Clarke2025PhD}.

\begin{figure}
    \centering
    \includegraphics[width=0.95\textwidth]{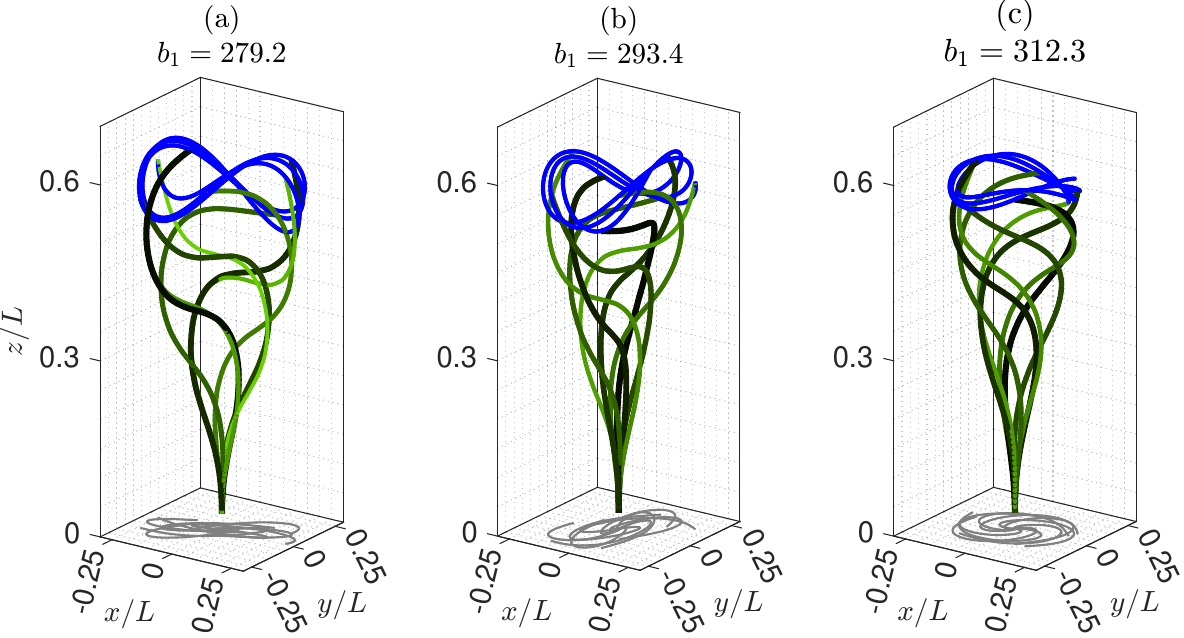}
    \caption{QP1 state throughout its existence for distributed surface flow model: (a) close to pitchfork bifurcation at $b_1=279$, (b) at an intermediate value of $b_1=293$, (c) close to Hopf bifurcation at $b_1=312$. The filament's colour gets darker in the increasing direction of the time. Filament's shadow is given in gray and its tip pathline is given in blue.}
    \label{fig:shapeevolution_qp1}
\end{figure}

We track QP1 in the bistable region as actuation increases, plotting its shape over time for the distributed surface flow model in Figure~\ref{fig:shapeevolution_qp1}. We see that QP1 carries the characteristics of both whirling and beating (see movie 1). At $b_1=279.2$, near the pitchfork bifurcation (see Figure~\ref{fig:shapeevolution_qp1}(a)), QP1 beats in a plane that is slowly whirling about the $z$-axis. Its tip pathline is similar to that of a beating. As actuation increases, whirling characteristics become stronger. Close to the Hopf bifurcation, at $b_1=312.3$, QP1 becomes a wobbly whirling state (see Figure~\ref{fig:shapeevolution_qp1}(c)), with the tip following a near-circular path. In the bistable region, the strength of the whirling mode increases with actuation, while that of the beating mode decreases.

\begin{figure}
    \centering
    \includegraphics[width=0.70\textwidth]{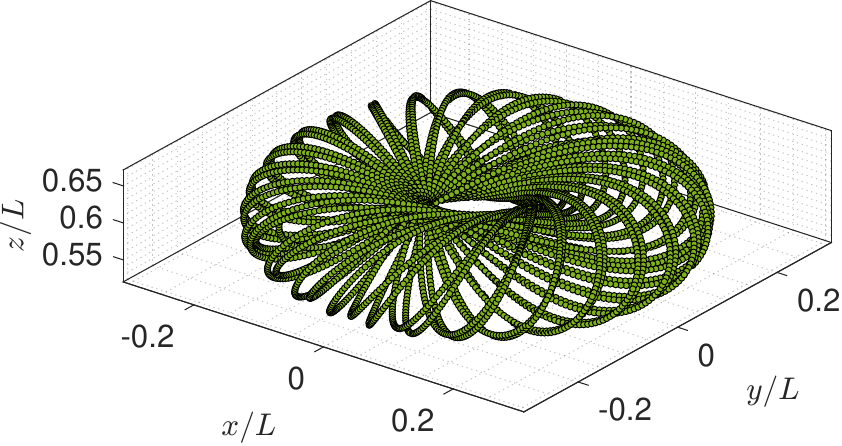}
    \caption{The tip pathline of QP1 state at $b_1=284$ for distributed surface flow model. The filament tip follows a path on a 2-torus.}
    \label{fig:qp1tip}
\end{figure}

\begin{figure}
    \centering
    \includegraphics[width=0.70\textwidth]{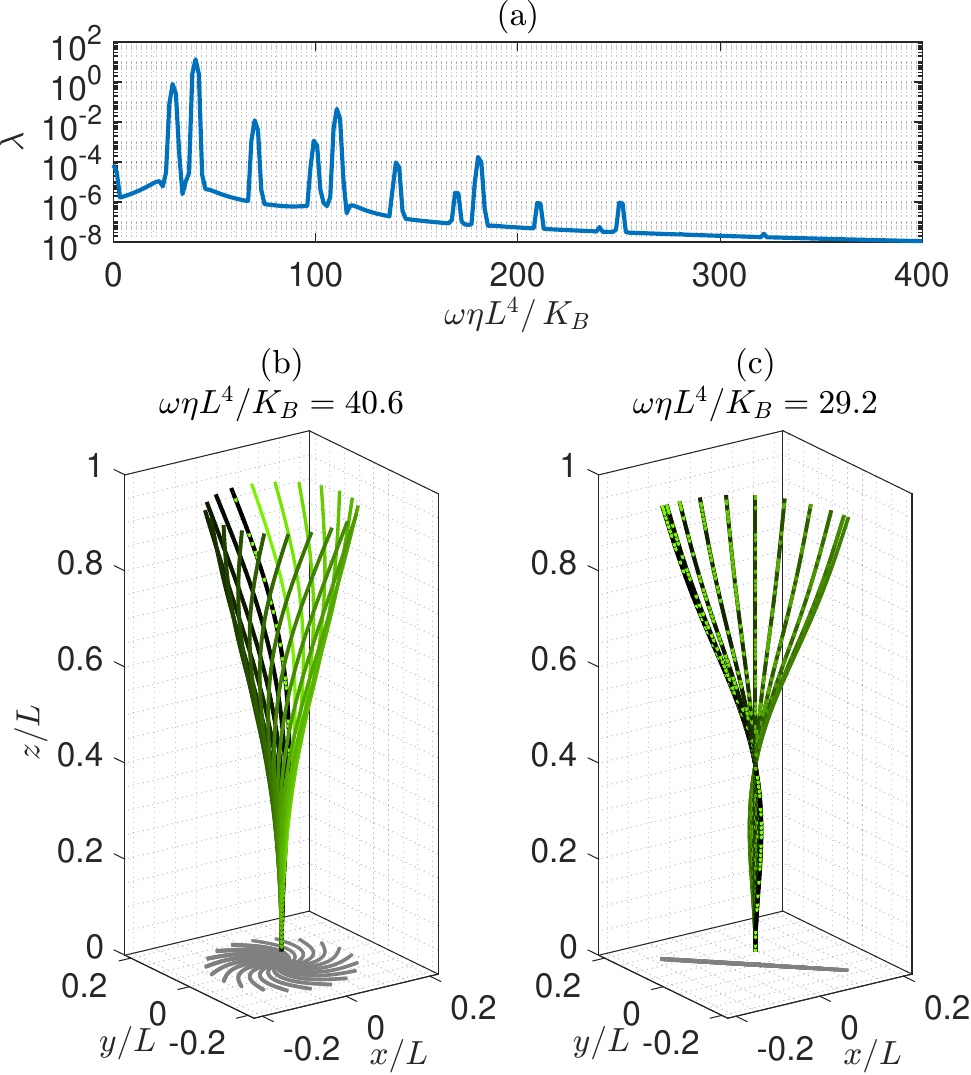}
    \caption{(a) Dimensionless SPOD frequency spectrum of QP1 at $b_1=742.5$ (close to the Hopf bifurcation) for tip-driven surface flow model. SPOD mode shapes associated with (b) the first and (c) the second dominant frequency. The filament shape in different phases of a mode are plotted in different colours. Filament's shadow is given in gray. The dimensionless frequency of whirling and beating states at $b_1=742.5$ are 41.6 and 34.2, respectively.}
    \label{fig:spod_qp1}
\end{figure}

Examining the tip pathline of the QP1 state in Figure~\ref{fig:qp1tip}, we see that it lies on a torus, which gives a qualitative indication of the quasiperiodicity of this solution. To examine this state quantitatively, we extract dominant frequencies and modes of QP1 using the SPOD approach described in \S\ref{sssec:spod} \citep{towne2018}. The SPOD spectrum (which is the leading eigenvalue given by SPOD at each frequency) of QP1 for the tip-driven surface flow model with $b_1=742.5$ is shown in Figure~\ref{fig:spod_qp1}(a). QP1 has two incommensurate dominant frequencies, consistent with its quasiperiodic nature. These frequencies, $40.6\,\eta L^4/K_B$ and $29.2\,\eta L^4/K_B$, are close to those of the whirling and beating states, $41.6\,\eta L^4/K_B$ and $34.2\,\eta L^4/K_B$, respectively, at the same actuation. The remaining peaks in the frequency spectrum reflect nonlinear interactions of the two dominant frequencies.

Figures~\ref{fig:spod_qp1}(b,c) show the two leading SPOD modes of QP1. At this actuation, which is close to the Hopf bifurcation where whirling becomes unstable, the leading SPOD mode (Figure~\ref{fig:spod_qp1}(b)) resembles a whirling state, which can be considered as the primary motion of QP1 at $b_1=742.5$. Figure~\ref{fig:spod_qp1}(c) shows the second mode, which is the leading mode associated with the second dominant frequency. We notice that this mode is planar and has a shorter wavelength than the eigenmode at the double Hopf bifurcation (Figure~\ref{fig:hhEigMode}(c)). This suggests that the second SPOD mode corresponds to a secondary buckling instability, which triggers the transition from whirling to beating. As the surface flow magnitude increases, the compression along the filament exceeds a critical load, leading to buckling with a wavenumber higher than that observed in the double Hopf bifurcation. This is consistent with the Floquet analysis, which also yielded higher wavenumber modes (see Figure \ref{fig:pbEigMode}).

\subsection{Instabilities and Dynamics at High Actuation}
\begin{figure}
    \centering
    \includegraphics[width=0.90\textwidth]{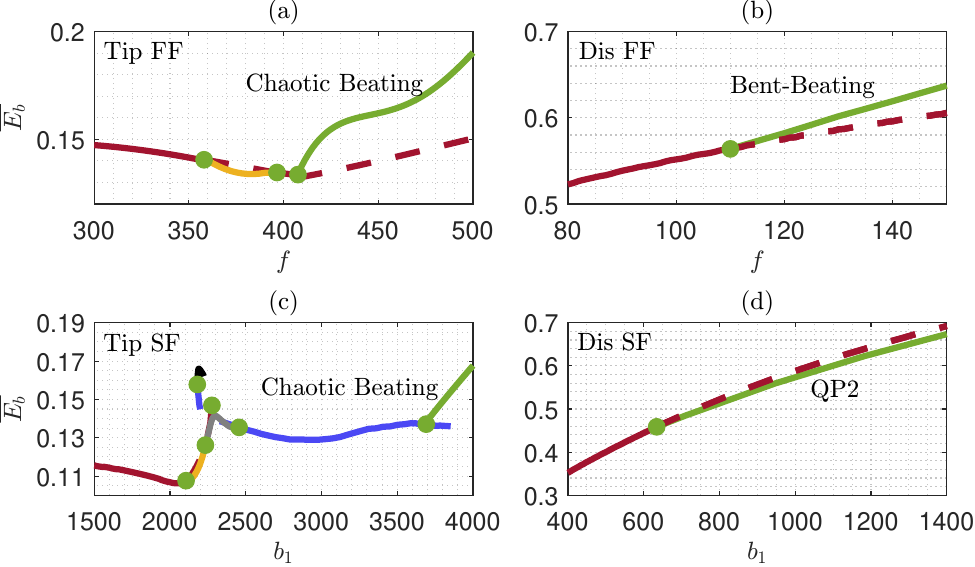}
    \caption{Bifurcation diagram for (a) tip-driven follower force, (b) distributed follower force, (c) tip-driven surface flow, and (d) distributed surface flow models at high actuation. Green full circles mark bifurcations. At the lower limit of each subplot a stable beating state is plotted. The final state exhibited by each model (which is model-dependent) is plotted in green.}
    \label{fig:bifdiag-High}
\end{figure}

\begin{figure}
    \centering
    \includegraphics[width=0.90\textwidth]{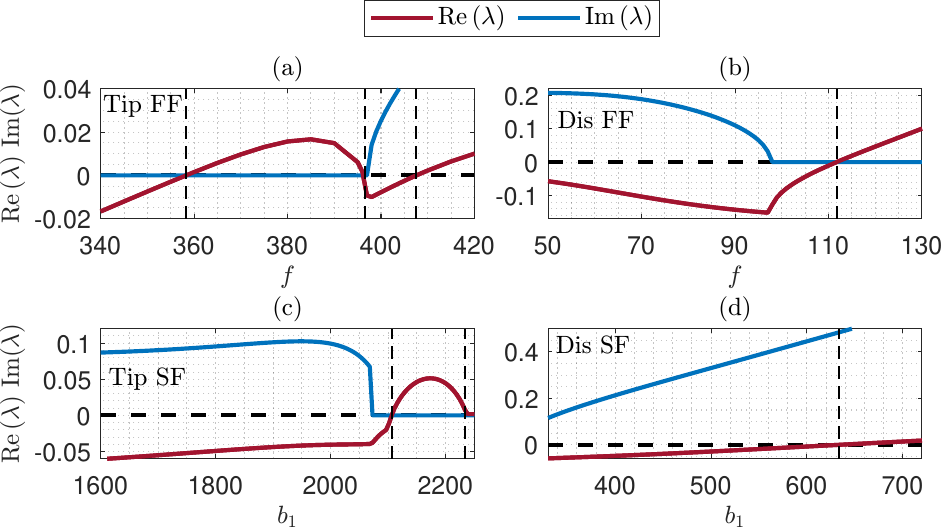}
    \caption{Leading eigenvalue of the beating state for (a) tip-driven follower force, (b) distributed follower force, (c) tip-driven surface flow, and (d) distributed surface flow models at high actuation. Real and imaginary parts of the leading eigenvalues are shown in red and blue, respectively. The bifurcations are marked with vertical dashed lines.}
    \label{fig:eigvals_beatingHigh}
\end{figure}

Up to moderate actuation, we observe all filament models exhibit the same dynamic states. As shown in \S\ref{ssec:stateslow}-\ref{ssec:wbqp1}, each model undergoes similar bifurcations that ultimately lead to a stable beating state. At higher actuations, however, beating becomes unstable and we encounter instabilities that yield different states for the different models.

The states encountered at high actuation for the different models are summarised in the bifurcation diagrams shown in Figure~\ref{fig:bifdiag-High}. In each, the stable beating state is plotted in red. The leading eigenvalues from the stability analysis of the beating state are shown in Figure~\ref{fig:eigvals_beatingHigh}. Each model undergoes distinct bifurcations and reaches a different state at the highest actuation values. These states are shown as green curves in Figure~\ref{fig:bifdiag-High}: a chaotic state for tip-driven models, a periodic bent-beating state for the distributed follower force model, and a quasiperiodic state, QP2, for the distributed surface flow model. We now describe the states exhibited by each model.

\subsubsection{Transition to Chaotic States for Tip-Driven Models}\label{ssec:tip-high}
\begin{figure}
    \centering
    \includegraphics[width=0.80\textwidth]{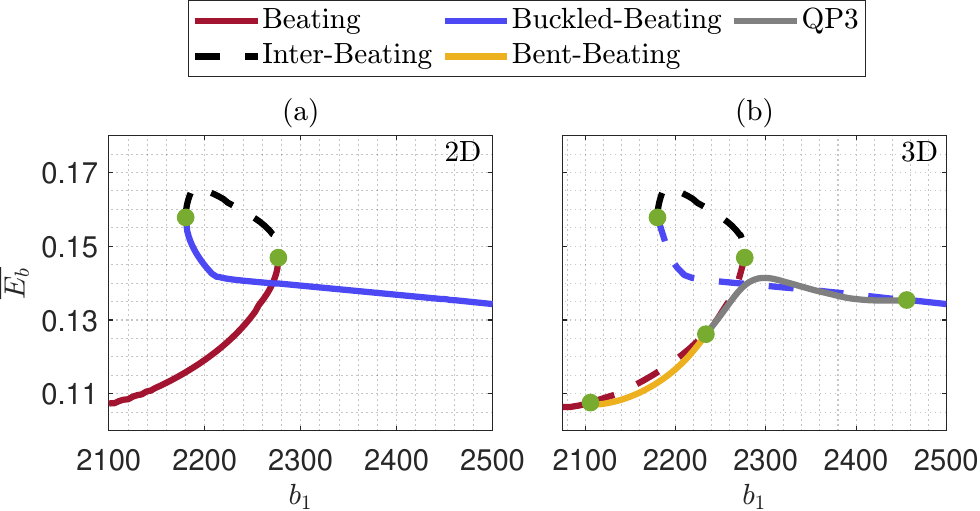}
    \caption{Close view of the transition regime of the bifurcation diagram in Figure~\ref{fig:bifdiag-High}(c) for tip-driven surface flow model. Green full circles mark bifurcations. (a) In 2D, we observe a saddle-node transition between three beating states: beating, inter-beating, and buckled-beating. (b) In addition to these, we observe two more bifurcations in the 3D model where bent-beating and QP3 emerge. After this transition, the state space of 2D and 3D dynamics are the same. At a higher surface flow value, we observe a transition to a planar chaotic state. We plot these states in Figures~\ref{fig:state-3beating}, \ref{fig:state-high}.}
    \label{fig:bif-sftipTrans}
\end{figure}

\begin{figure}
    \centering
    \includegraphics[width=0.95\textwidth]{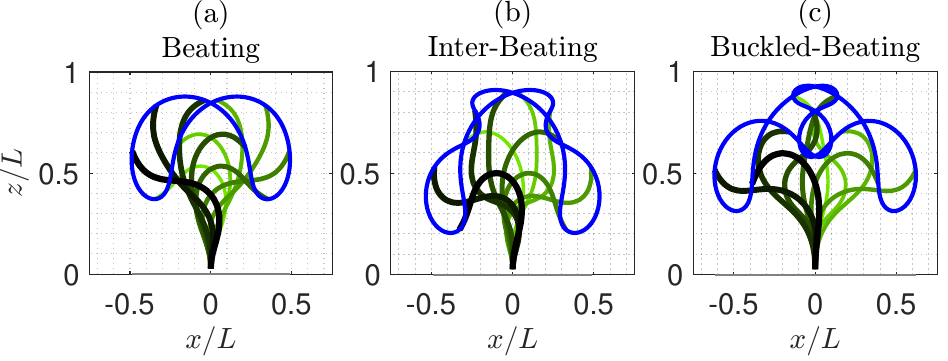}
    \caption{(a) Beating, (b) inter-beating, (c) buckled-beating states coexisting in the transition regime at $b_1=2220$, obtained using the tip-driven surface flow model. All three of the states are planar, hence exist in both 2D and 3D. The tip pathline of the filament is plotted in blue. The filament colour gets darker as time increases. We provide a video corresponding to this figure in movie 2.}
    \label{fig:state-3beating}
\end{figure}

\begin{figure}
    \centering
    \includegraphics[width=0.90\textwidth]{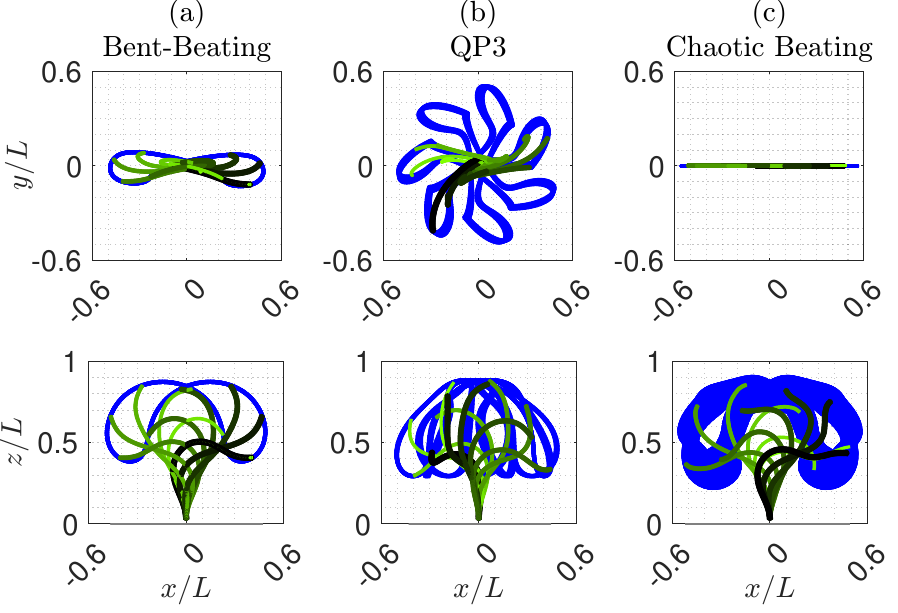}
    \caption{Filament states at high actuation: (a) bent-beating, (b) QP3, (c) chaotic beating. The plotted states are obtained using the tip-driven surface flow model at $b_1=2220$, $b_1=2270$, and $b_1=4020$. The tip pathline of the filament is plotted in blue. The filament colour gets darker in the increasing direction of time. We provide a video description of this figure in movie 2.}
    \label{fig:state-high}
\end{figure}

For the tip-driven models, we observe a myriad of states and transitions that emerge at high actuation, consistent with previous studies on the tip-driven follower force model \citep{ling2018,clarke2024}. In the discussion that follows, we focus exclusively on the tip-driven surface flow model, thoroughly mapping its instabilities and bifurcations, as this case has not been reported previously.

To begin, we explore the dynamics when filament motion is restricted to a plane. Here, we find three states connected by two saddle-node bifurcations, shown in Figure~\ref{fig:bif-sftipTrans}(a). The three coexisting beating states plotted in Figure~\ref{fig:state-3beating} are (a) beating, (b) inter-beating, and (c) buckled-beating. Beating and buckled-beating are bistable and the unstable inter-beating solution forms the branch that joins these two states.

In our discussion, we set time such that a beat cycle starts and ends ($t=0$ and $t=T$) when the filament tip attains its maximum amplitude in the $-x$-direction; the filament tip is on the $z$-axis at $t=T/4$ and $t=3T/4$, and attains its maximum amplitude in the $x$-direction at $t=T/2$. In the beating state, the filament remains curved throughout the beat cycle (see Figure~\ref{fig:state-3beating}(a)). In inter-beating and buckled-beating, the filament is nearly upright at $t=T/4$ and $t=3T/4$. This is especially noticeable in buckled-beating, Figure~\ref{fig:state-3beating}(c). At these instances, the nearly straight filament experiences compression that triggers dynamics reminiscent of a higher wavenumber buckling event. This is reflected in the downward motion of the filament tip. Nonlinear effects follow this buckling transient, and the filament continues along the rest of its beat path. Comparing Figure~\ref{fig:state-3beating}(b) with (c), we observe that buckling occurs along a shorter section of the filament in inter-beating and begins after the tip crosses $x=0$. Tracking the unstable inter-beating branch shows that the buckled section of the filament decreases in length with increasing actuation.

When we allow for fully-3D filament motion, more complex bifurcations (Figure~\ref{fig:bif-sftipTrans}(b)) are observed. Evaluating the stability of these 2D states in 3D, we find that beating becomes unstable before its annihilation, and buckled-beating emerges as an unstable state. Furthermore, we discover two new states: bent-beating and QP3 as visualised in Figure~\ref{fig:state-high}(a,b).

First, we observe a supercritical pitchfork bifurcation at $b_1=2106$ and the creation of bent-beating. Bent-beating is similar to beating, but the tip of the filament curves out of the beating plane (see Figure~\ref{fig:state-high}(a)). We find that the leading unstable mode is orthogonal to the plane of the beating state, leading to 3D dynamics. The bent-beating branch (orange in Figure~\ref{fig:bif-sftipTrans}(b)) is stable throughout its existence. By increasing the surface flow magnitude, we see that the out-of-plane tip deflection peaks at $b_1=2172$ and then declines, with bent-beating eventually merging back into beating at $b_1=2242$.

A second unstable mode of beating emerges as the surface flow magnitude increases and produces a supercritical pitchfork bifurcation at $b_1=2234$. As this occurs before the annihilation of bent-beating, beating becomes unstable to two modes and continues to be unstable throughout. QP3 emerges from this bifurcation (grey in Figure~\ref{fig:bif-sftipTrans}(b)), and is destroyed at $b_1=2456$ in yet another supercritical pitchfork bifurcation with the buckled-beating state.  Plotted in Figure~\ref{fig:state-high}(b), QP3, albeit quasiperiodic, is different from QP1 and QP2 (discussed in \S\ref{sssec:dis-high}). QP3 exhibits dynamics that combine out-of-plane tipping, as seen with bent-beating, with the periodic buckling encountered in the 2D periodic states in Figure~\ref{fig:state-3beating}. The buckling mechanism is weak around $b_1=2234$ (near beating) and becomes stronger as the actuation increases toward $b_1=2456$ (near the buckled-beating state). Tipping-buckling coupling causes sudden changes in the beating plane, seen in the top-view in Figure~\ref{fig:state-high}(b).

We note that the bifurcation patterns are simpler for the tip-driven follower force model. When beating becomes unstable at $f=358$, we observe only the bent-beating state, compared to the many states observed in the tip-driven surface flow model (see Figure~\ref{fig:bifdiag-High}(a,c)). The leading eigenvalue is real in Figure~\ref{fig:eigvals_beatingHigh}(a) (as it is in the case of the tip-driven surface flow model), and due to the symmetry breaking in the out-of-plane direction, we observe a pitchfork bifurcation. We suspect that greater variety of states exhibited by the surface flow model is related to the stress peak at the filament tip that arises in this model and, presumably, produces shorter wavelength buckling events.

The final state observed in tip-driven models at the highest values of actuation we consider is a chaotic state (green in Figure~\ref{fig:bifdiag-High}(a,c)). This state arises as multiple buckled-beating eigenmodes become unstable over a small actuation range. It exhibits high-wavenumber instabilities near the filament tip (see movie 2) and exists in both 2D and 3D (Figure~\ref{fig:state-high}(c)).

\subsubsection{Dynamics of Distributed Actuation Models at High Actuation}\label{sssec:dis-high}
We find that when the actuation profile is uniform along the filament, the bifurcation patterns at high actuation are very different from those of tip-driven models. In both of the distributed actuation models, we find a single attracting state emerges after beating becomes unstable (see Figures~\ref{fig:bifdiag-High}(b,d)). As in \S\ref{ssec:tip-high}, we start by classifying the dynamics for the surface flow model.

\begin{figure}
    \centering
    \includegraphics[width=0.95\textwidth]{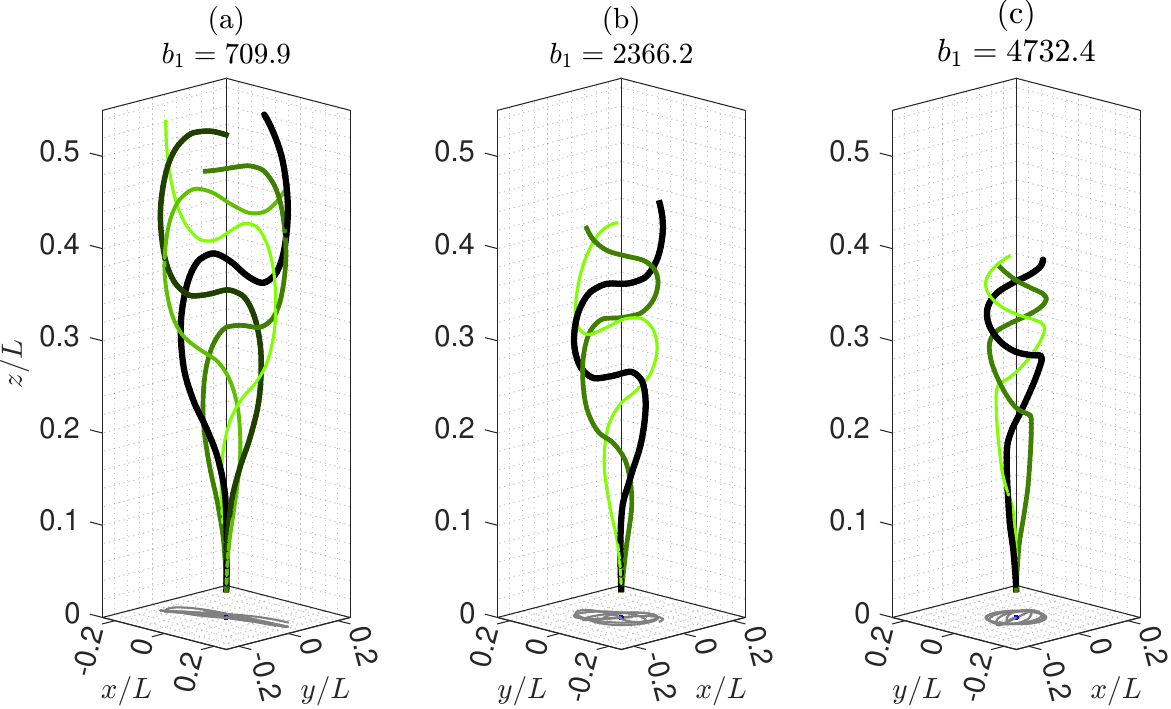}
    \caption{QP2 state with increasing actuation for the distributed surface flow model, at (a) $b_1=709.9$, (b) $b_1=2366$, and (c) $b_1=4732$. The filament's colour gets darker as time passes. Filament's shadow is given in gray. We provide a video description of this figure in movie 3.}
    \label{fig:shapeevolution_qp2}
\end{figure}

Figure~\ref{fig:eigvals_beatingHigh}(d) shows the leading eigenvalue (a complex conjugate pair) at high actuation, revealing a supercritical Hopf bifurcation for the distributed surface flow model. We identify a new quasiperiodic state, QP2. Figure~\ref{fig:shapeevolution_qp2} tracks QP2 along its branch as the actuation increases. Near $b_1=634$, QP2 exhibits in-plane beating combined with an out-of-plane motion, evident from the nearly planar shadow of the filament with some out-of-plane components (see Figure~\ref{fig:shapeevolution_qp2}(a)). Since this instability is absent in 2D, we conclude that the out-of-plane instability is the main mechanism driving the creation of QP2.

As actuation increases, the shadow of the filament changes from planar to circular, indicating a balance between the primary modes of QP2. The filament forms a rotating helix-like shape (Figure~\ref{fig:shapeevolution_qp2}(c)). Presumably as a result of the surface flow, the filament is inclined to move toward the base and generate further compression, resulting in a more shortened shape overall (compare Figures~\ref{fig:shapeevolution_qp2}(a) and (c)).

\begin{figure}
    \centering
    \includegraphics[width=0.80\textwidth]{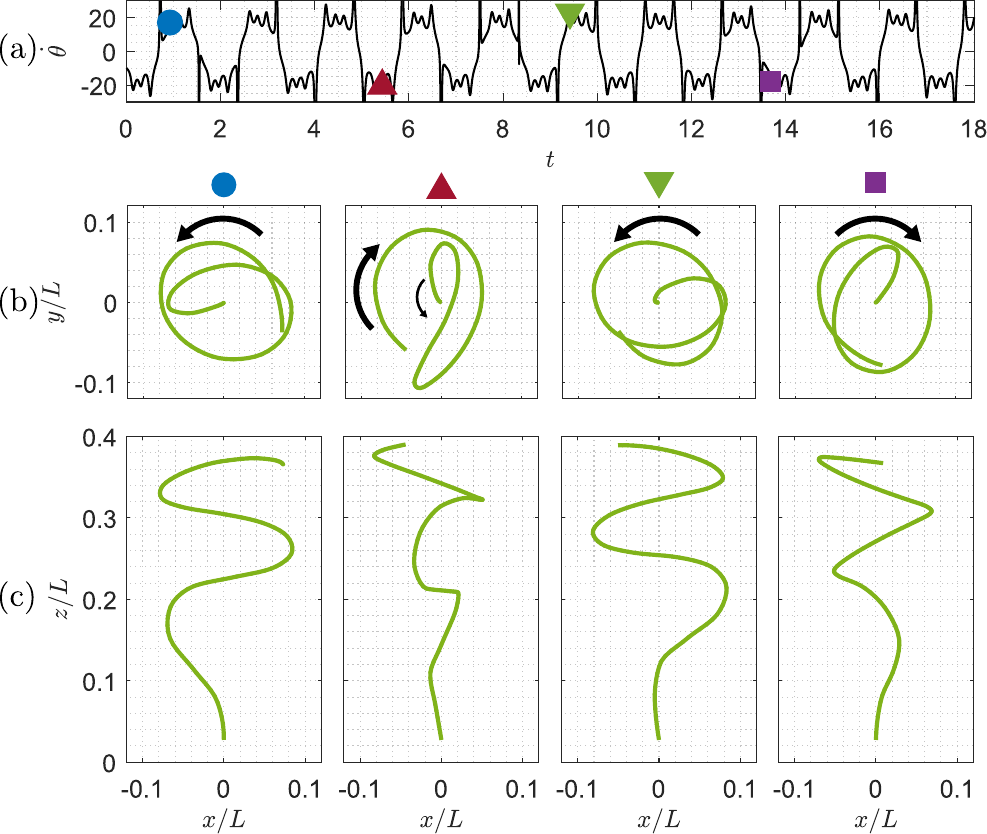}
    \caption{(a) Angular speed of filament tip about the $z$-axis, for QP2 state at $b_1=3310$ for the distributed squirming model. The filament shapes in the four instances marked are plotted from (b) top view and (c) side view. The rotation direction of the filament are indicated by black arrows.}
    \label{fig:thtdot-qp2}
\end{figure}

A notable feature of QP2 is the handedness of the filament coils. Coils form at the base and advect towards the tip. Given the handedness of the helix-like shape and the overall rotation of the filament, it is inclined to move toward its base. This further increases compression at the base and triggers buckling. This creates kinks that also travel towards the tip and lead to the reversal of the coiling direction. Due to reversal, the handedness above and below a kink are opposite of each other.

We track the angular position, $\theta$, of the filament tip extracted from its $x$- and $y$-coordinates and from it compute the angular speed, $\Dot{\theta}$. The angular speed of the filament tip in Figure~\ref{fig:thtdot-qp2}(a) shows reversals linked to changes in coiling direction at the base. We mark four instances and plot the filament shape at those instances together with the rotation directions in Figures~\ref{fig:thtdot-qp2}(b,c). The time stamps are in sequential order and are separated by five tip reversals. In the first instance (blue circle), the filament has a positive angular speed with left-handed coils without kinks. In the second instance, the angular speed of the tip is negative while the base rotates anticlockwise. The filament has left-handed coils near the base and right-handed coils close to the free end, with a visible kink in-between. The third instance returns to fully left-handed coils (see movie 3(c)). These transient reversals of handedness are unique to QP2 and the distributed surface flow model. Increasing actuation further reveals that QP2 is robust, remaining the only stable state that we find. We also track the unstable beating state and find no additional instabilities at higher actuation.

Although both distributed actuation models exhibit a single stable state at high actuation, the characteristics of the transition differ. The distributed follower force model undergoes a supercritical pitchfork bifurcation (see Figure~\ref{fig:eigvals_beatingHigh}(b)), leading to bent-beating (see Figure~\ref{fig:state-high}(a), discussed in \S\ref{ssec:tip-high}). Unlike in the tip-driven follower force model, bent-beating is robust to increasing actuation. These attracting states in both distributed models suggest that the uniform actuation profile stabilises the filament against high-wavenumber instabilities. This is, presumably, due to the peak stress being near the base rather than the free end.

\section{Conclusions}\label{sec:Conclusions}
In this paper, we carried out a bifurcation and stability analysis of various active filament models, discovering new states and identifying transition mechanisms. We considered a single active filament clamped at its base in an unbounded domain driven by two actuation models (follower force and surface flow) and two actuation profiles (tip-driven and distributed). We described the instability mechanisms that lead to new states and transitions.

Examining the filament dynamics starting from zero actuation, we found that all models exhibited similar initial states and transitions. This indicates that the qualitative features of the dynamics and their bifurcations are independent of actuation type or profile. The motion of the surrounding fluid had a stabilising effect in surface flow models, delaying the primary and secondary instabilities. Distributed actuation yielded a stress profile that has its maximum value at the filament base, leading to unstable mode shapes that are narrower at the tip. Bistability between whirling and beating was observed in all models except the tip-driven surface flow model. The existence of QP1 is a generic feature; however, its stability is sensitive to model details. We found that the transition regime involving whirling, beating, and QP1 is linked to secondary buckling, with compression playing the primary role in the bifurcations.

At high actuation, the filament dynamics differ due to actuation profile. Tip-driven models exhibit different periodic or chaotic solutions, while distributed models exhibit states robust to increasing actuation. In particular, we found an interesting coiled state (QP2) in the distributed surface flow model which featured a rotating helix-like shape that changes handedness. We presume that the differences in the states exhibited by the models is linked to the stress distributions that we explored in more detail at lower actuation. There, distributed models had a stress profile that decreased toward the free end. Tip-driven models had the opposite profile with the stress increasing toward the free end. We suspect that this profile triggers high-wavenumber instabilities at high actuation. We note that the torque supplied by the internal stresses after buckling (i.e. $\that\times\bLambda$ in (\ref{eq:contBalance})) in tip-driven models is generally expected to be larger at the tip than that in distributed models. Additionally, given the clamped boundary condition at the base, the tangent vector rotation is also less restricted at the filament tip. Thus, the larger internal stress observed around the tip in tip-driven models coupled with the ease with which the filament tip can rotate, likely results in large filament deformations ultimately producing chaotic beating dynamics.

The filament models we investigated are biologically relevant for understanding microtubule-generated flows. Microtubules have a bending rigidity of about 22 pN$\mu$m$^2$ and can exceed 40 $\mu$m in length \citep{gittes1993}. Their motion is driven by dynein motors exerting forces up to 6 pN, which can induce buckling and drive dynamic states \citep{shingyoji1998,ling2018}. These correspond to follower force magnitudes up to $f=440$ or equivalently surface flow magnitudes up to around $b_1=1530$ in our dimensionless units. These values fall within the parameter range studied. In fact, similar states to the ones observed in this study also exist in nature, such as ciliary beating in the fallopian tube \citep{xiao2017}, quasiperiodic motion of filopodia in HeLa cells \citep{chen2014}, or helical buckling of actin-rich filaments in glioma cells \citep{schell2012}.

\begin{figure}
    \centering
    \includegraphics[width=0.80\textwidth]{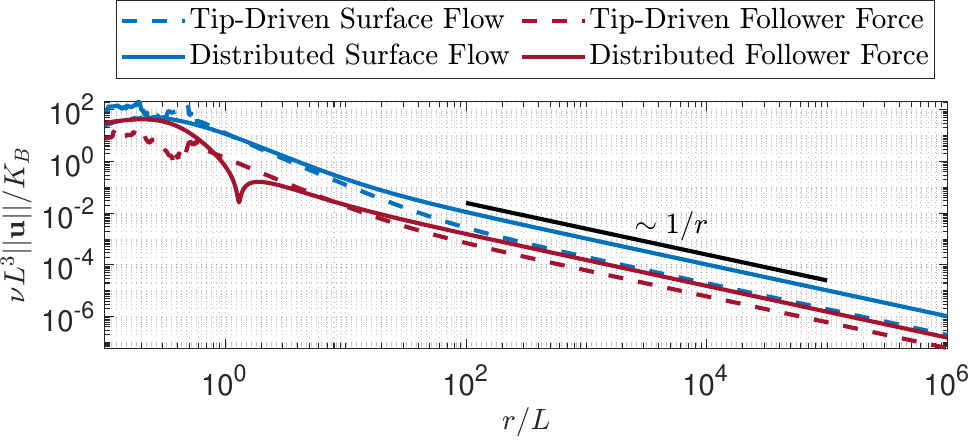}
    \caption{The period-averaged flow magnitude of the beating state for different active filament models. The chosen actuation values are $f=396.5$ and $f=109.7$ for follower force models and $b_1=3691$ and $b_1=634.1$ for surface flow models, which are at the bifurcation point (marked in Figure~\ref{fig:eigvals_beatingHigh}) leading to the annihilation of the beating state. Regardless of the model, the flow decays as $1/r$.}
    \label{fig:flowdecay}
\end{figure}

While our study has focused on the filament dynamics, it is important to note that the resulting flow fields for the follower force and surface flow models differ significantly, even in cases where the filament dynamics appear similar. For example, in the upright state, the follower force is balanced by internal stresses, resulting in no net fluid flow. In contrast, the surface flow model generates considerable fluid motion. After buckling, filament motion in the follower force model will produce a fluid flow, however, the flow generated by the surface flow model remains considerably stronger. As an example, Figure~\ref{fig:flowdecay} shows the period-averaged flow magnitude of the beating state at actuation values where beating annihilates: $f=396.5$ and $f=109.7$ for the follower force models, and $b_1=3691$ and $b_1=634.1$ for the surface flow models. For tip-driven models, these states correspond to the onset of chaotic dynamics. All active filament models exhibit a $1/r$ decay in the far field, with the distributed surface flow model producing the strongest flow.

Although the differences in the flow field do not largely affect the qualitative dynamics and bifurcations at low actuation, this may not be the case where multiple filaments interact. Indeed, when multiple filaments are present, their hydrodynamic interactions are able to alter the bifurcations that are observed and, most notably change the initial double Hopf bifurcation to a pitchfork bifurcation where the filaments are found to bend in the same direction \citep{stein2021}. In contrast, with follower forces alone, only coordinated beating or whirling has been seen to emerge \citep{westwood2021coordinated,su2024}. These fluid-structure interactions have been shown to contribute to the emergence of vortical flows involved in rapid mixing and transport within cells \citep{dutta2024}. Our future work will focus on examining these and related bifurcations that occur in biofilament clusters, as well as bifurcations and instabilities in models driven by active shearing \citep{oriola2017nonlinear} and incorporating structural features such as anisotropy \citep{clarke2025structural}.

\appendix
\section{Derivation of the Slip Velocity Correction}\label{app:squirmMat}
\subsection{Single Force-Free Squirmer}
The squirmer is a spherical swimming body that imposes instantaneous velocity boundary conditions to the surrounding fluid due to effective surface deformations \citep{lighthill1952,blake1971}. A force-free, neutral squirmer swimming in the $\ehat$-direction applies the force quadruple \citep{ishikawa2006,delmotte2015}, $\boldsymbol{H}=-4/3\pi\eta a^3B_1\ehat$, on the fluid and generates a Stokes flow that satisfies
\begin{align}\label{eq:stokesEqSingle}
    \boldsymbol{0}& =-\bnabla p+\eta\bnabla^2\boldsymbol{u}+\boldsymbol{H}\,\bnabla^2\boldsymbol{\delta}\left(\br\right), \\
    \boldsymbol{0}& =\bnabla \cdot \boldsymbol{u},
\end{align}
where $\br$ is the position relative to the squirmer centre, $r=\lVert\br\rVert$, and $\rhat=\br/r$. The solution to this system reads
\begin{equation}
    \boldsymbol{u}(\br)=\dfrac{1}{4\pi\eta r^3}\left(\idvec-3\,\rhat\rhat\right)\boldsymbol{H}.
\end{equation}

We obtain the squirmer-induced velocity, $\boldsymbol{u}^\prime$, of a nearby particle with hydrodynamic radius $a$ using Fax{\'e}n's relations \citep{kim2013},
\begin{equation} \label{eq:squirmv}
    \boldsymbol{u}^\prime=\left(\idvec+\dfrac{a^2}{6}\bnabla^2\right)\,\boldsymbol{u}(\br).
\end{equation}
Since, in this case, the contribution of the Laplacian term is zero, the induced velocity is simply $\boldsymbol{u}(\br)$ and given by \eqref{eq:squirmv}.  On the other hand, the self-induced velocity of a squirmer is $2/3B_1\ehat$.

\subsection{Hydrodynamically Coupled Squirmers}
For an $N$-squirmer system, we have
\begin{equation}
    \boldsymbol{0}=-\bnabla p+\eta\bnabla^2\boldsymbol{u}+\sum\limits_i\boldsymbol{H}_i\,\bnabla^2\boldsymbol{\delta}\left(\boldsymbol{x}-\bY_i\right)
\end{equation}
where $\boldsymbol{H}_i=4/3\pi\eta a^3B_1\that_i$. Here, we choose the swimming direction $\ehat =-\that$ such that each squirmer tries to move toward the base of the filament. Then, the flow induced on segment $j$ by segment $i$ ($i\neq j$) is
\begin{equation}\label{eq:sqmat1}
    \boldsymbol{u}^\prime\left(\br_{ij}=\bY_i-\bY_j\right)=\dfrac{1}{4\pi\eta r_{ij}^3}\left(\idvec-3\,\rhat_{ij}\rhat_{ij}\right)\boldsymbol{H}_j=\boldsymbol{S}_{ij}\boldsymbol{H}_j
\end{equation}
for a squirming interaction matrix $\boldsymbol{S}_{ij}\in\mathbb{R}^{3\times3}$. We remark that $i,j$ are not matrix indices, but denote segment indices. Similarly, self-interactions can be expressed as
\begin{equation}\label{eq:sqmat2}
    \boldsymbol{u}^\prime\left(\br_{ii}\right)=-\dfrac{1}{2\pi\eta a^3}\idvec\,\boldsymbol{H}_i=\boldsymbol{S}_{ii}\boldsymbol{H}_i.
\end{equation}

Using \eqref{eq:sqmat1} and \eqref{eq:sqmat2}, we can group all $\boldsymbol{H}_i$ under a general swimming vector $\boldsymbol{H}=\left[\boldsymbol{H}_1^T,\cdots,\boldsymbol{H}_N^T\right]^T\in\mathbb{R}^{3N\times1}$ and a general swimming matrix $\mathcal{S}\in\mathbb{R}^{3N\times3N}$ with block entries $\boldsymbol{S}_{ij}$. Then, the translational and angular velocities due to the segment surface flows can be expressed as,
\begin{equation}\label{eq:squirmingMatEq}
    \begin{bmatrix}
        \boldsymbol{V}^\prime\\
        \boldsymbol{\Omega}^\prime
    \end{bmatrix}=\begin{bmatrix}
        \mathcal{S}&\boldsymbol{0}\\\boldsymbol{0}&\boldsymbol{0}
    \end{bmatrix}\begin{bmatrix}
        \boldsymbol{H} \\\boldsymbol{0}
    \end{bmatrix}.
\end{equation}

\section{The Effective Lie Algebra Element}\label{app:effLieAlgEl}
It is sufficient to track the tangent angle along the filament to describe the 2D filament dynamics. However, in 3D, a different approach is required to describe the evolution of a local frame at each point along the filament centreline. In our numerical framework, following \cite{clarke2024}, we use unit quaternions to describe local rotations of the coordinate frame and effective Lie algebra elements for the state space representation of the dynamics.

Given the equilibrium quaternion, which describes the upright filament,
\begin{equation}
    \boldsymbol{q}_\text{eq}=\begin{bmatrix}
        \dfrac{1}{\sqrt{2}},0,-\dfrac{1}{\sqrt{2}},0
    \end{bmatrix}\qquad\text{with}\qquad\boldsymbol{q}^*_\text{eq}=\begin{bmatrix}
        \dfrac{1}{\sqrt{2}},0,\dfrac{1}{\sqrt{2}},0
    \end{bmatrix},
\end{equation}
the effective Lie algebra element associated with a quaternion $\boldsymbol{q}$ is defined as
\begin{equation}
    \bU = 2\arccos{\left(p_0\right)}\hat{\boldsymbol{p}},
\end{equation}
where $\boldsymbol{q}\bullet\boldsymbol{q}^*_\text{eq}=\begin{bmatrix}
    p_0,\bar{\boldsymbol{p}}
\end{bmatrix}$ and $\hat{\boldsymbol{p}}=\dfrac{\bar{\boldsymbol{p}}}{\lVert\bar{\boldsymbol{p}}\rVert}$.

\subsection{Connection to the Position Vector}\label{ConnctntoPosVec}
Given an effective Lie algebra element $\bU(s_k)$ associated with the $k$-th filament segment, the corresponding quaternion is given by
\begin{equation}
    \boldsymbol{q}_k=\exp{\left(\bU(s_k)\right)}\bullet\boldsymbol{q}_\text{eq},
\end{equation}
where $\exp{\left(\bU\right)}=\begin{bmatrix}
    \cos{\dfrac{\lVert\bU\rVert}{2}},\sin{\dfrac{\lVert\bU\rVert}{2}}\dfrac{\bU}{\lVert\bU\rVert}
\end{bmatrix}$. Then, the centre position of the $k$-th segment is found recursively as
\begin{equation}
    \mathbf{X}_k=\mathbf{X}_{k-1}+\dfrac{\Delta L}{2}\left(\mathbf{R}\left(\boldsymbol{q}_k\right)\ehat_1+\mathbf{R}\left(\boldsymbol{q}_{k-1}\right)\ehat_1\right),
\end{equation}
where the action of the rotation matrix on an arbitrary vector $\mathbf{v}$ is defined as
\begin{equation*}
    \begin{bmatrix}
        0,\mathbf{R}\left(\boldsymbol{q}\right)\mathbf{v}
    \end{bmatrix}=\boldsymbol{q}\bullet\begin{bmatrix}
        0,\mathbf{v}
    \end{bmatrix}\bullet \boldsymbol{q}^*.
\end{equation*}

Then, the position of a segment centre is
\begin{equation}
\begin{aligned}
    \bX_k&=\bX_1+\dfrac{\Delta L}{2}\left(\mathbf{R}\left(\boldsymbol{q}_1\right)+\mathbf{R}\left(\boldsymbol{q}_k\right)+2\sum_2^{k-1}\mathbf{R}\left(\boldsymbol{q}_i\right)\right)\hat{\mathbf{e}}_1\\
    &=\bX_1+\dfrac{\Delta L}{2}\vect\left(\boldsymbol{q}_1\bullet\begin{bmatrix}0,\hat{\mathbf{e}}_1\end{bmatrix}\bullet\boldsymbol{q}^*_1+\boldsymbol{q}_k\bullet\begin{bmatrix}0,\hat{\mathbf{e}}_1\end{bmatrix}\bullet\boldsymbol{q}^*_k+2\sum_2^{k-1}\boldsymbol{q}_i\bullet\begin{bmatrix}0,\hat{\mathbf{e}}_1\end{bmatrix}\bullet\boldsymbol{q}^*_i\right),
\end{aligned}
\end{equation}
where $\vect$ operator takes the vector part of the quaternion. Notice that each quaternion multiplication term corresponds to the tangent vector of the respective filament segment. We investigate some special cases of the effective Lie algebra element to show its connection to the tangent vectors, and hence the position and geometry of the filament.

\subsection{Special Cases of Effective Lie Algebra Element}\label{SpecialCases}
\subsubsection{Case I:}\label{case1}
If the effective Lie algebra element of a segment is of the form
\begin{equation}
    \bU(s,t)=\begin{bmatrix}U_1(s,t),0,0\end{bmatrix},
\end{equation}
then the associated tangent vector of a segment is
\begin{equation}
\begin{aligned}
    \hat{\mathbf{t}}(s,t)&=0\,\hat{\mathbf{e}}_1-\sin\lVert\bU(s,t)\rVert\,\hat{\mathbf{e}}_2+\cos\lVert\bU(s,t)\rVert\,\hat{\mathbf{e}}_3\\
    &=0\,\hat{\mathbf{e}}_1-\sin\lvert U_1(s,t)\rvert\,\hat{\mathbf{e}}_2+\cos\lvert U_1(s,t)\rvert\,\hat{\mathbf{e}}_3.
\end{aligned}
\end{equation}
This form of effective Lie algebra element relates to planar motion in the $yz$- plane.

\subsubsection{Case II:}\label{case2}
If the effective Lie algebra element of a segment is of the form
\begin{equation}
    \bU(s,t)=\begin{bmatrix}0,U_2(s,t),0\end{bmatrix},
\end{equation}
then the associated tangent vector of a segment is
\begin{equation}
\begin{aligned}
    \hat{\mathbf{t}}(s,t)&=\sin\lVert\bU(s,t)\rVert\,\hat{\mathbf{e}}_1+0\,\hat{\mathbf{e}}_2+\cos\lVert\bU(s,t)\rVert\,\hat{\mathbf{e}}_3\\
    &=\sin\lvert U_2(s,t)\rvert\,\hat{\mathbf{e}}_1+0\,\hat{\mathbf{e}}_2+\cos\lvert U_2(s,t)\rvert\,\hat{\mathbf{e}}_3.
\end{aligned}
\end{equation}
This form of effective Lie algebra element relates to planar motion in the $xz$- plane.

\subsubsection{Case III:}\label{case3}
A natural generalization of the previous cases is their sum, that is an effective Lie algebra element of a segment is of the form
\begin{equation}
    \bU(s,t)=\begin{bmatrix}U_1(s,t),U_2(s,t),0\end{bmatrix}.
\end{equation}
The associated tangent vector of a segment is
\begin{equation}
    \hat{\mathbf{t}}(s,t)=\dfrac{U_2(s,t)}{\lVert\bU(s,t)\rVert}\sin\lVert\bU(s,t)\rVert\,\hat{\mathbf{e}}_1-\dfrac{U_1(s,t)}{\lVert\bU(s,t)\rVert}\sin\lVert\bU(s,t)\rVert\,\hat{\mathbf{e}}_2+\cos\lVert\bU(s,t)\rVert\,\hat{\mathbf{e}}_3.
\end{equation}
In the case where $\lVert\bU(s,t)\rVert$ is time independent, such that $U_1^2(s,t)+U_2^2(s,t)=c(s)$ for a function $c(s)$, the tangent vector follows a circular path restricted to a $z=\text{constant}$ plane. This is a model for a whirling-like motion of the filament.

On the other hand, if the norm of the Lie algebra element changes but ${U_2}/{U_1}=\text{constant}$ (or ${U_1}/{U_2}=\text{constant}$, whichever is well-defined), the motion is restricted to a plane. The angle the plane makes with the $x$-axis is $\theta=-\arctan({U_1}/{U_2})$.

\subsubsection{Case IV:}
The effective Lie algebra element in its most general form is
\begin{equation}
    \bU(s,t)=\begin{bmatrix}U_1(s,t),U_2(s,t),U_3(s,t)\end{bmatrix}.
\end{equation}
The resulting tangent vector is
\begin{equation}\label{eq:UtoT}
\begin{aligned}
    \hat{\mathbf{t}}(s,t)=&\left[\dfrac{U_1}{\lVert\bU\rVert}\dfrac{U_3\left(1-\cos\lVert\bU\rVert\right)}{\lVert\bU\rVert}+\dfrac{U_2}{\lVert\bU\rVert}\sin\lVert\bU\rVert\right]\,\hat{\mathbf{e}}_1\\
    &\left[\dfrac{U_2}{\lVert\bU\rVert}\dfrac{U_3\left(1-\cos\lVert\bU\rVert\right)}{\lVert\bU\rVert}-\dfrac{U_1}{\lVert\bU\rVert}\sin\lVert\bU\rVert\right]\,\hat{\mathbf{e}}_2\\
    &\left[\dfrac{U_3}{\lVert\bU\rVert}\dfrac{U_3\left(1-\cos\lVert\bU\rVert\right)}{\lVert\bU\rVert}+\cos\lVert\bU\rVert\right]\,\hat{\mathbf{e}}_3.
\end{aligned}
\end{equation}

\subsection{Connection to Periodic States}
As shown in Appendix~\ref{SpecialCases}, certain types of the effective Lie algebra element induce certain restrictions on the filament dynamics. Here, we generalise these results as two examples of filament states.

\subsubsection{Decomposition of Beating State}
Beating is a planar state, which can be described as a time-periodic bending of the filament. Since it is essentially a two-dimensional state, a segment's tangent vector can be expressed by only one time-dependent variable,
\begin{equation}
    \mathbf{\hat{{t}}}(s,t)=\sin{\left(\phi(s,t)\right)}\cos{\theta}\,\hat{\mathbf{e}}_1- \sin{\left(\phi(s,t)\right)}\sin{\theta}\,\hat{\mathbf{e}}_2+\cos{\left(\phi(s,t)\right)}\,\hat{\mathbf{e}}_3,
\end{equation}
where $\theta$ determines the beating plane and $\phi(s,t)$ is the beat function. From Appendix~\ref{case3}, we take $U_1(s,t)=\phi(s,t)\sin\theta$ and $U_2(s,t)=\phi(s,t)\cos\theta$. Then, the associated effective Lie algebra element is
\begin{equation}
\begin{aligned}
    \bU(s,t)&=\begin{bmatrix}
        \phi(s,t)\sin{\theta},\phi(s,t)\cos{\theta},0
    \end{bmatrix}\\
    &=\sin{\theta}\begin{bmatrix}
        \phi(s,t),0,0
    \end{bmatrix}+\cos{\theta}\begin{bmatrix}
        0,\phi(s,t),0
    \end{bmatrix}.
\end{aligned}
\end{equation}

Notice that we do not restrict neither the filament shape nor its time-dependency. Here, we only assume (owing to periodicity) that $\phi(s,t)=\phi(s,t+T)$ where $T$ is the beating period. This decomposition also demonstrates that the summation of two beating signals (with unit modulus) corresponds to rotations about the $z$-axis.

\subsubsection{Decomposition of Whirling State}
The tangent vector of a whirling filament can be described without loss of generality as
\begin{equation}
    \that(s,t)=Ae^{i\omega t}\sin\psi(s)\,\hat{\mathbf{e}}_1-Ae^{i\omega t}\sin\psi(s)\,\hat{\mathbf{e}}_2+\cos\psi(s)\,\hat{\mathbf{e}}_3+cc.
\end{equation}
for some arbitrary $A\in\mathbb{C}$ with $\lvert A\rvert=1$. Here, $\tan\psi(s)$ is the time-independent angle that the tangent vector makes from the meridional point of view. From Appendix~\ref{case3}, we take $U_1^2(s,t)^2+U_2^2(s,t)^2=\psi^2(s)$, and obtain $U_1(s,t)=iAe^{i\omega t}\psi(s)$ and $U_2(s,t)=Ae^{i\omega t}\psi(s)$. Then, any whirling Lie algebra element can be decomposed as
\begin{equation}
\begin{aligned}
    \bU(s,t)&=\begin{bmatrix}
        iA\psi(s)e^{i\omega t},A\psi(s)e^{i\omega t},0
    \end{bmatrix}+cc.\\
    &=\left(iA\begin{bmatrix}
        \psi(s),0,0
    \end{bmatrix}+A\begin{bmatrix}
        0,\psi(s),0
    \end{bmatrix}\right)e^{i\omega t}+cc..
\end{aligned}
\end{equation}
Each of these effective Lie algebra elements describes planar filament dynamics - one on the $yz$-plane and another on the $xz$-plane. They are identical in shape, up to a $\pi/2$ rotation about the $z$-axis. The periodic time dependence is separated from space; therefore, the filament shape is, in a sense, ``frozen in time'' and whirling is a solid body rotation.

However, another interpretation of this result is that these effective Lie algebra elements individually describe a beating filament. The beating has the frequency $\omega/(2\pi)$ and the filament has a specific shape described by $\phi(s)$. The coefficients associated with each signal have the same magnitude, but are orthogonal to each other in the complex plane. Therefore, a $\pi/2$-delayed combination of two beating signals creates whirling. Furthermore, the sum of any two beating signals is not necessarily a whirling state, in that whirling requires time and space to be separated in the beating function.

\subsection{Use in the Bisection Method}\label{app:bisec}
Given $\bU_\mathrm{w}(s,t)=[U_{1,\mathrm{w}}(s,t),U_{2,\mathrm{w}}(s,t),0]$ for whirling, we have
\begin{equation}\label{eq:biseq2}
    U_{1,\mathrm{w}}^2(s)+U_{2,\mathrm{w}}^2(s)=c(s)
\end{equation}
for some $c(s)\in\mathbb{R}$. Similarly, given $\bU_\mathrm{b}(s,t)=[U_{1,\mathrm{b}}(s,t),U_{2,\mathrm{b}}(s,t),0]$ for beating, we have
\begin{equation}\label{eq:biseq3}
    \dfrac{U_{1,\mathrm{b}}(s)}{U_{2,\mathrm{b}}(s)}=c
\end{equation}
for some constant $c\in\mathbb{R}$. After a sufficiently long time, a solution satisfies either \eqref{eq:biseq2} or \eqref{eq:biseq3}, which we use as convergence criteria.

\bibliography{ref}

\end{document}